\documentclass[amsmath,amssymb,nofootinbib,prd,onecolumn,notitlepage,superscriptaddress]{revtex4-1}
\pdfoutput=1 
\usepackage{graphicx}
\usepackage{caption}
\usepackage{subcaption}
\usepackage{epsfig}
\usepackage{rotate}
\usepackage{amsmath}
\usepackage{amssymb}
\usepackage{amsfonts}
\usepackage{enumerate}
\usepackage{afterpage}
\usepackage{xcolor}
\usepackage{natbib}
\usepackage{bm}
\usepackage{mathtools}
\usepackage{hyperref}
\usepackage{float}
\usepackage[export]{adjustbox}
\usepackage[alsoload=astro]{siunitx} 
\usepackage[normalem]{ulem}

\allowdisplaybreaks 
\DeclarePairedDelimiter\evaluat{.}{\rvert}
\reDeclarePairedDelimiterInnerWrapper\evaluat{star}{%
    \mathopen{}\mathclose\bgroup #1\hskip -\nulldelimiterspace \relax
    #2\aftergroup\egroup #3%
}

\usepackage{blkarray}

\renewcommand{\mathbf}{\bm}
\newcommand{\n}{\bm {n}}
\renewcommand{\k}{\bm {k}}

\newcommand{\ka}{\bm {k}_1}
\newcommand{\kb}{\bm {k}_2}
\newcommand{\kc}{\bm {k}_3}
\newcommand{\ko}{\mathcal{K}}

\newcommand{\p}{\partial}
\newcommand{\cH}{\mathcal{H}}
\newcommand{\fnl}{f_\text{NL}}

\newcommand{\diff}{\,\mathrm{d}}
\newcommand*\mean[1]{\bar{#1}}
\DeclareSIUnit\erg{erg}

\newcommand{\Q}{\mathcal{Q}}

\def\be{\begin{equation}}
\def\ee{\end{equation}}
\def\bea{\begin{eqnarray}}
\def\eea{\end{eqnarray}}
\def\i{\mathrm{i\,}}

\begin{document}

\title{Multipoles of the relativistic galaxy bispectrum}

\author{Eline M. de Weerd}
\affiliation{School of Physics \& Astronomy, Queen Mary University of London, London E1 4NS, UK}

\author{Chris Clarkson}
\affiliation{School of Physics \& Astronomy, Queen Mary University of London, London E1 4NS, UK}
\affiliation{Department of Physics \& Astronomy, University of the Western Cape, Cape Town 7535, South Africa}
\affiliation{Department of Mathematics \& Applied Mathematics, University of Cape Town, Cape Town 7701, South Africa}

\author{Sheean Jolicoeur}
\affiliation{Department of Physics \& Astronomy, University of the Western Cape, Cape Town 7535, South Africa}

\author{Roy Maartens}
\affiliation{Department of Physics \& Astronomy, University of the Western Cape, Cape Town 7535, South Africa}
\affiliation{Institute of Cosmology \& Gravitation, University of Portsmouth, Portsmouth PO1 3FX, UK}

\author{Obinna Umeh}
\affiliation{Institute of Cosmology \& Gravitation, University of Portsmouth, Portsmouth PO1 3FX, UK}

\date{\today}

\begin{abstract}

Above the equality scale the galaxy bispectrum will be  a key probe for measuring primordial non-Gaussianity which can help differentiate between different inflationary models and other theories of the early universe. On these scales a variety of relativistic effects come into play once the galaxy number-count fluctuation is projected onto our past lightcone. By decomposing the Fourier-space bispectrum into invariant multipoles about the observer's line of sight we examine in detail how the relativistic effects contribute to these. We show how to perform this decomposition analytically, which is significantly faster for subsequent computations.  While all multipoles receive a contribution from the relativistic part, odd multipoles arising from the imaginary part of the bispectrum have no Newtonian contribution, making the odd multipoles a smoking gun for a relativistic signature in the bispectrum for single tracers.  The dipole and the octopole are significant on equality scales and above where the Newtonian approximation breaks down. This breakdown is further signified by the fact that the even multipoles receive a significant correction on very large scales.

\end{abstract}
\maketitle

\section{Introduction}
The bispectrum will play a key role in future galaxy surveys as an  important probe of large-scale structure and for measuring primordial 
non-Gaussianity and galaxy bias~\cite{Jeong:2009vd,Baldauf:2010vn,Celoria:2018euj}. It can help discriminate between different inflationary models and other theories of the early universe, and contains information that is complementary and additional to what is contained in the power spectrum. 
On super-equality scales, a variety of relativistic effects come into play once the galaxy number-count fluctuation is projected onto the past light cone. In the density contrast up to second order, relativistic effects arise from observing on the past lightcone, and they include all redshift, volume and lensing distortions and couplings between these. In Poisson gauge, these effects can be attributed to velocities (Doppler), gravitational potentials (Sachs-Wolfe, integrated SW, time delay) and lensing magnification and shear. In addition, there are corrections arising from a GR definition of galaxy bias~\cite{Bertacca:2014wga}. These effects generate corrections to the Newtonian approximation at order $\mathcal{O}(\cH/k)$ and higher. Non-Gaussianity generated by these relativistic projection effects could closely mimic the signature of \(\fnl\) on large scales which gives a correction in the halo bias $\mathcal{O}((\cH/k)^2)$, indicating the importance of precisely including all $\mathcal{O}(\cH/k)$ and higher effects in theoretical modelling. So far, a variety of relativistic effects in the galaxy Fourier bispectrum has been taken into account, see 
\cite{Umeh:2016nuh,Jolicoeur:2017nyt,Jolicoeur:2017eyi,Jolicoeur:2018blf,Clarkson:2018dwn,Maartens:2019yhx} under the assumption of the plane parallel approximation, and neglecting integrated effects. Other groups are working on this from different angles and approaches, for example by a spherical-Fourier formalism~\cite{Bertacca:2017dzm}, and calculating the angular galaxy bispectrum~\cite{DiDio:2016gpd,DiDio:2018unb}. Crucially, we have shown that the relativistic part should be detectable in a survey like Euclid without resorting to the multi-tracer technique, which is needed for the power spectrum~\cite{Maartens:2019yhx} . 

Once an observable like the galaxy number-count fluctuation is projected onto the past lightcone the orientation of the triangle in the Fourier space bispectrum becomes important. 
Analogously to how the Legendre multipole expansion is used for power spectrum analysis, one can expand the galaxy bispectrum in spherical harmonics, thus isolating the different invariant multipoles with respect to the observer's line of sight $\bm n$. We use the full spherical harmonics for the bispectrum rather than the Legendre polynomial expansion usually adopted for the power spectrum because of the azimuthal degrees of freedom associated with the orientation of the triangle with respect to the line of sight direction vector in Fourier space. In the power spectrum limit, there is only one angular degree of freedom after ensemble averaging. For the bispectrum, we have one angular and one azimuthal degree of freedom which when expanded in spherical harmonics leads to $(2\ell + 1)$ independent harmonics for each multipole value $\ell$. 

This has been done for the Newtonian bispectrum, which generates non-zero multipoles only for even 
\(\ell\) (up to \(\ell = 8\)) due to redshift-space distortions~\cite{Scoccimarro_1999,Nan:2017oaq}. Contrary to the Newtonian bispectrum, the relativistic galaxy bispectrum generates non-zero multipoles for both even and odd \(\ell\) up to \(\ell = 8\) and \(m = 6\) where the odd multipoles are induced by the general relativistic effects only. This means that these multipoles are a crucial signature of relativistic projection effects. We provide, for the first time, a multipole decomposition of the Fourier space galaxy bispectrum with relativistic effects included. Additionally we show that the coefficients of this expansion can be worked out analytically. We provide an exact analytic formula for this multipole expansion of the galaxy bispectrum. Previously, we examined for the first time the dipole of the galaxy bispectrum in detail, showing that its amplitude can be more than 10\% of that of the monopole even at equality scales~\cite{Clarkson:2018dwn}. 
In order to eliminate possible biases when analysing large scale structure data, it is important to include the relativistic effects. In addition to this, a variety of the effects that appear in the bispectrum are relativistic effects that have not been measured elsewhere and hence are interesting to study. By analysing the non-zero multipoles of the galaxy bispectrum both for a Euclid-like galaxy survey, and for an SKA-like HI intensity mapping survey, we show the behaviour of the higher multipoles and their corrections to the Newtonian bispectrum. In follow-up work, we are investigating possibilities of detecting the higher multipoles of the bispectrum. See for example~\cite{Maartens:2019yhx} for detection prospects of the leading order relativistic effects; the dipole is expected to have the strongest GR signature. 

The paper is organised as follows. We introduce the relativistic Fourier space bispectrum in section~\ref{sec:relbisp}, and present the multipole expansion of the relativistic bispectrum in section~\ref{sec:extrmulti}. An analysis of the multipoles can be found in section~\ref{sec:anal}. Finally, we summarise our conclusions in section~\ref{sec:concl}.

\section{The relativistic bispectrum}\label{sec:relbisp}

In Fourier space, the observed galaxy bispectrum \(B_g\) at a fixed redshift \(z\) is given by~\cite{Jolicoeur:2017nyt,Jolicoeur:2017eyi}
\begin{equation}
	\langle \Delta_g(z,\ka) \Delta_g(z,\kb) \Delta_g(z,\kc) \rangle = (2\pi)^3 B_g(z,\ka,\kb,\kc) \delta^D(\ka+\kb+\kc),
\end{equation}
where \(\Delta_g(z, \ka)\) is the number count contrast at redshift \(z\) (see~\cite{Jolicoeur:2017nyt} for the full expression). Here we work in the Poisson gauge; note that \(\Delta_g = \delta_g + \text{RSD}+\text{GR}\) projection effects, where the RSD term is the Kaiser RSD up to second order, which is part of the Newtonian approximation. Since redshift is fixed, in what follows we drop redshift dependence for brevity. Furthermore, since the observed direction \(\n\) is fixed in what follows, the plane-parallel approximation is necessarily assumed. Then, at tree level, and for Gaussian initial conditions, the following combinations of terms contribute, 
\begin{equation}
\langle \Delta_g(\ka)\Delta_g(\kb)\Delta_g(\kc) \rangle = \frac{1}{2} \langle \Delta_g^{(1)}(\ka)\Delta_g^{(1)}(\kb)\Delta_g^{(2)}(\kc)\rangle \text{ + 2 cyclic permutations.}
\end{equation}
Using Wick's theorem, this gives an expression for the galaxy bispectrum ~\cite{Jolicoeur:2017nyt}
\begin{equation}
B_g(\ka, \kb,\kc)= \ko^{(1)}(\ka) \ko^{(1)}(\kb) \ko^{(2)}(\ka,\kb,\kc)P(\ka) P(\kb)\text{ + 2 cyclic permutations},
\end{equation}
where \(P\) is the power spectrum of \(\delta_\mathrm{T}^{(1)}\), the first order dark matter density contrast in the total-matter gauge, which corresponds to an Eulerian frame. The first order kernel can be split into a Newtonian and a relativistic part as~\cite{Jeong_2012}
\begin{equation}
\ko^{(1)} = \ko^{(1)}_\mathrm{N} + \ko^{(1)}_\mathrm{GR}, \qquad \ko^{(1)}_\mathrm{N}= b_1 + f \mu^2, \qquad \ko^{(1)}_\mathrm{GR} = \i \mu \frac{\gamma_1}{k} + \frac{\gamma_2}{k^2}, 
\end{equation}
with \(\mu = \hat\k \cdot \n \) (\(\hat\k = \k/k\)), \(b_1\) is the first-order Eulerian galaxy bias coefficient, \(f\) is the linear growth rate of matter perturbations, and redshift-dependent coefficients \(\gamma_i\) are~\cite{Jeong_2012},
\begin{align}\label{eq:gamma1def}
\frac{\gamma_1}{\cH} &= f \left[ b_e - 2 \Q - \frac{2(1-\Q)}{\chi \cH} - \frac{\cH'}{\cH^2}\right], \\
\frac{\gamma_2}{\cH^2} &= f(3-b_e) + \frac{3}{2} \Omega_m \left[ 2+ b_e - f - 4\Q - \frac{2(1-\Q)}{\chi\cH} - \frac{\cH'}{\cH^2} \right].\label{eq:gamma2def}
\end{align}
In equations~\eqref{eq:gamma1def}~and~\eqref{eq:gamma2def}, \(\cH\) is the conformal Hubble rate \((\ln a)'\), where a prime denotes a derivative with respect to conformal time; \(b_e\) and \(\Q\) are the galaxy evolution and magnification biases respectively, \(\chi\) is the line-of-sight comoving distance and \(\Omega_m = \Omega_{m0} (1+z) H_0^2/\cH^2\) is the matter density parameter. At first order, the gauge-independent GR definition of galaxy bias is made in the common comoving frame of galaxies and matter, 
\begin{equation}\label{eq:comovinggalaxybiasdef}
	\delta_{g\mathrm{C}}^{(1)} = b_1 \delta_\mathrm{C}^{(1)} =  b_1 \delta_\mathrm{T}^{(1)}\,,
\end{equation}
where subscript C is for the comoving gauge and T is for total matter gauge, which is a gauge corresponding to standard Newtonian perturbation theory. The bias relation in Poisson gauge is then obtained by transforming~\eqref{eq:comovinggalaxybiasdef} to Poisson gauge~\cite{Bertacca:2014wga,Jolicoeur:2017eyi}:
\begin{equation}\label{eq:deltareln}
	\delta_g^{(1)} = \delta_{g\mathrm{C}}^{(1)} + (3 - b_e)\cH v^{(1)} = b_1 \delta_\mathrm{T}^{(1)} + (3-b_e)\cH v^{(1)},
\end{equation}
where \(v^{(1)}\) is the velocity potential. Since \(v^{(1)} =f \cH \delta_T^{(1)}/k^2\), the last term on the right of equation~\eqref{eq:deltareln} leads to the \(f(3-b_e)\) term in \(\gamma_2/\cH^2\), equation~\eqref{eq:gamma2def}.

Similarly to the first order kernel, the second order kernel can be split into a Newtonian and a relativistic part. The second order part of the Newtonian kernel is well studied and is given as~\cite{Bernardeau_2002,Karagiannis_2018,Scoccimarro_1999,Verde_1998}
\begin{align}\label{newt2o}
	\ko_\mathrm{N}^{(2)}(\ka, \kb, \kc) &= b_1 F_2(\ka,\kb) + b_2 + f \mu_3^2 G_2(\ka,\kb) + f Z_2(\ka,\kb) + b_{s^2} S_2(\ka,\kb),
\end{align}
where \(\mu_i = \hat\k_i \cdot \n \), \( b_2\) is the second-order Eulerian bias parameter, and \(b_{s^2} \) is the tidal bias. \(F_2\) and \(G_2\) are the Fourier-space Eulerian kernels for second-order density contrast and velocity respectively~\cite{Jolicoeur:2017nyt,Villa:2015ppa}; 
\begin{align}
F_2(\ka,\kb) &= 1 + \frac{F}{D^2} + \left(\hat{\k}_1 \cdot \hat{\k}_2 \right)\left( \frac{k_1}{k_2} + \frac{k_2}{k_1}\right) + \left( 1-\frac{F}{D^2}\right) \left(\hat{\k}_1 \cdot \hat{\k}_2 \right)^2,\nonumber \\
G_2(\ka,\kb) &= \frac{F'}{D D'} + \left(\hat{\k}_1 \cdot \hat{\k}_2\right)\left(\frac{k_1}{k_2} + \frac{k_2}{k_1} \right) + \left( 2 - \frac{F'}{D D'} \right) \left(\hat{\k}_1 \cdot \hat{\k}_2\right)^2,
\end{align}
where \(F\) is a second-order growth factor, which is given by the growing mode solution of,
\begin{equation}
F'' + \cH F' - \frac{3}{2}\frac{H_0^2 \Omega_{m0}}{a} F = \frac{3}{2}\frac{H_0^2 \Omega_{m0}}{a} D^2.
\end{equation}
 In an Einstein-de Sitter background, \(F= 3 D^2 / 7\), which is a very good approximation for \(\Lambda\)CDM which we use here. The second-order RSD part of the Newtonian kernel is comprised of \(G_2\) above and the kernel \(Z_2\)~\cite{Verde_1998,Scoccimarro_1999},
\begin{equation}
	Z_2(\ka,\kb) = f \frac{\mu_1 \mu_2}{k_1 k_2}\left( \mu_1 k_1 + \mu_2 k_2 \right)^2 + \frac{b_1 }{k_1 k_2} \left[ \left( \mu_1^2 + \mu_2^2 \right)k_1 k_2 + \mu_1 \mu_2 \left( k_1^2 + k_2^2 \right) \right]\,.
\end{equation}

 Finally, \(S_2(\ka,\kb)\) is the kernel for the tidal bias,
\begin{equation}
	S_2(\ka,\kb) = (\hat{\k}_1 \cdot \hat{\k}_2)^2 - \frac{1}{3}\,.
\end{equation}
The Newtonian bias model is 
\begin{equation}\label{eq:newtbiasmodel}
	\delta_{g\mathrm{T}}^{(2)} = b_1 \delta_\mathrm{T}^{(2)} + b_2 \left[\delta_\mathrm{T}^{(1)}\right]^2 + b_{s^2} s^2\,,
\end{equation}
where \( s^2 = s_{ij}s^{ij}\), and \( s_{ij} = \Phi_{,ij} - \delta_{ij}\nabla^2 \Phi /3\) . 

The relativistic part of the second order kernel was first derived in~\cite{Umeh:2016nuh} in the simplest case and extended in~\cite{Jolicoeur:2017nyt,Jolicoeur:2017eyi,Jolicoeur:2018blf}. Neglecting sub-dominant vector and tensor contributions, we have 
\begin{align}\label{eq:soGRkernel}
\ko_\mathrm{GR}^{(2)}(\ka,\kb,\kc) &= \frac{1}{k_1^2 k_2^2} \left\{ \vphantom{\frac{k_1^2 k_2^2}{k_3}} \beta_1 + E_2(\ka,\kb,\kc) \beta_2 
+\i (\mu_1k_1+\mu_2k_2)\beta_3
+ \i \mu_3 k_3\left[ \beta_{4} + E_2(\ka, \kb, \kc)\beta_{5}\right] \right. \nonumber \\
&\left. +\frac{k_1^2 k_2^2}{k_3^2} \left[ F_2(\ka,\kb) \beta_{6} + G_2(\ka,\kb) \beta_{7} \right] + \left( \mu_1 k_1 \mu_2 k_2 \right)\beta_{8} + \mu_3^2 k_3^2 \left(\beta_{9} + E_2(\ka,\kb,\kc) \beta_{10} \right) \right. \nonumber \\
&\left. + \left(\ka \cdot \kb \right)\beta_{11} + \left( k_1^2 + k_2^2 \right) \beta_{12} + \left( \mu_1^2 k_1^2 + \mu_2^2 k_2^2 \right) \beta_{13} \right. \nonumber \\
&\left. + \i \left[\vphantom{\frac{k_1^2 k_2^2}{k_3}} \left( \mu_1 k_1^3 + \mu_2 k_2^3 \right)\beta_{14} + \left( \mu_1 k_1 + \mu_2 k_2 \right) \left(\ka\cdot\kb \right) \beta_{15} + k_1 k_2 \left(\mu_1 k_2 + \mu_2 k_1 \right) \beta_{16} \right. \right. \nonumber \\
&\left. \left. + \left( \mu_1^3 k_1^3 + \mu_2^3 k_2^3 \right)\beta_{17} + \mu_1 \mu_2 k_1 k_2 \left( \mu_1 k_1 + \mu_2 k_2 \right) \beta_{18} + \mu_3 \frac{k_1^2 k_2^2}{k_3} G_2(\ka,\kb) \beta_{19} \right] \right\}.
\end{align}
We have collected terms according to the overall powers of $k$ involved.
The \(\beta_i\) here are redshift- and bias-dependent coefficients, given in full in appendix~\ref{sec:betacoeffappendix}, which updates  expressions in previous papers.
We have defined the kernel \(E_2\) which scales as \(k^0\) (like \(F_2\), \(G_2\), and \(Z_2\) do), 
\begin{equation}
E_2(\ka,\kb,\kc) = \frac{k_1^2 k_2^2}{k_3^4} \left[ 3 + 2 \left( \hat{\k}_1 \cdot\hat{\k}_2 \right) \left( \frac{k_1}{k_2} + \frac{k_2}{k_1} \right) + \left( \hat{\k}_1 \cdot \hat{\k}_2 \right)^2 \right],
\end{equation}
which incorporates some of the relativistic dynamical corrections to the intrinsic second-order terms.

At second order, the GR bias model, which corrects the Newtonian bias model~\eqref{eq:newtbiasmodel} is given by~\cite{Umeh:2019qyd},
\begin{equation}
	\delta_{g\mathrm{T}}^{(2)} = b_1 \delta_\mathrm{T}^{(2)} + b_2 \left[ \delta_\mathrm{T}^{(1)} \right]^2 + b_{s^2} s^2 + \delta_{\mathrm{C,GR}}^{(2)}\,,
\end{equation}
where the last term maintains gauge invariance on ultra-large scales, and is given by (using \( \delta_\mathrm{C}^{(1)} =  \delta_\mathrm{T}^{(1)}\) )
\begin{equation}\label{eq:grcorrdelta}
	\delta_{\mathrm{C,GR}}^{(2)} = 2 \cH^2 (3 \Omega_m + 2 f) \left[ \delta_\mathrm{T}^{(1)} \nabla^{-2} \delta_\mathrm{T}^{(1)} - \frac{1}{4} \partial_i \nabla^{-2} \delta_\mathrm{T}^{(1)} \partial^i \nabla^{-2} \delta_\mathrm{T}^{(1)} \right]\,.
\end{equation}
The GR correction~\eqref{eq:grcorrdelta} to the Newtonian bias model is contained in the GR kernel~\eqref{eq:soGRkernel}. Then, we also need to transform \(\delta_{g\mathrm{T}}^{(2)}\) to the Poisson	 gauge \( \delta_g^{(2)} \), the expression for this is given in~\cite{Jolicoeur:2017nyt}, 
\begin{align}
\delta_{g}^{(2)}  &= \delta_{g{\rm T} }^{(2)} + (3-b_e) \cH v^{(2)} +\Big[ (b_e-3) \cH' + b_e' \cH + (b_e-3)^2 \cH^2 \Big]  \big[v^{(1)}\big]^2 + (b_e-3)\cH   v^{(1)}  v^{(1)\prime} \nonumber\\
& +2(3-b_e)\cH  v^{(1)} \delta_{g{\rm T}}^{(1)}  - 2  v^{(1)} \delta_{g{\rm T}}^{(1)\prime} + 3 \left( b_e-3\right) \cH v^{(1)}\Phi^{(1)} \,.\label{dgpt}
\end{align}
All of the terms after $\delta_{g\mathrm{T}}^{(2)}$ on the right of equation~\eqref{dgpt} scale as $(\cH/k)^n \left[ \delta^{(1)}_\mathrm{T} \right]^2$, where $n = 2,4$. Therefore they are omitted in the Newtonian approximation. These GR correction terms maintain gauge-independence on ultra-large scales, and they are included in the GR kernel~\eqref{eq:soGRkernel}.

\section{Extracting the multipoles}\label{sec:extrmulti}

\begin{figure}[H]
	\centering
	\includegraphics[width=0.6
	\linewidth]{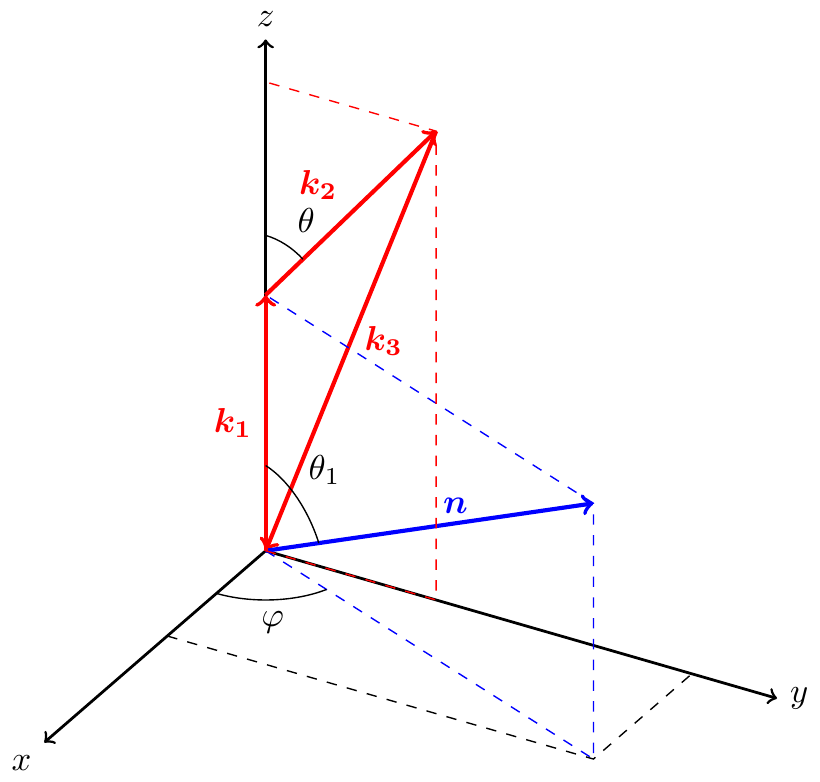}
	\caption{Overview of the relevant vectors and angles for the Fourier-space bispectrum. \label{fig:geometry_overview} }
\end{figure}

Our goal is to extract the spherical harmonic multipoles of $B_g$ with respect to the observer's line of sight. That is, for a fixed line of sight and triangle shape, the rotation of the plane of the triangle about $\n$ generates invariant moments, the sum of which add up to the full bispectrum. 
 This means that  
\be
B_g =\sum_{\ell m} B_{\ell m}Y_{\ell m}(\n)\,,
\ee
where we follow~\cite{Scoccimarro_1999,Nan:2017oaq} in our choice of decomposition of the bispectrum (an alternative basis can be found in~\cite{Sugiyama_2018}).
To define the $B_{\ell m}$ we need to define an orientation for the $Y_{\ell m}$ to give the polar and azimuthal angles over which to integrate. We choose a coordinate basis for the vectors that span the triangle as follows: 
\begin{eqnarray}
	&&\k_1 = (0,0,k_1)\, \\
	&&\k_2 = (0, k_2 \sin\theta, k_2 \cos\theta)\,, \\
	&&\k_3 = (0, - k_2 \sin\theta, - k_1 - k_2\cos\theta )\,,\\
	&&\bm{n} = (\sin\theta_1 \cos\varphi, \sin\theta_1 \sin\varphi,\cos\theta_1)\,.
\end{eqnarray}
That is, we fix $\k_1$ along the $z$-axis, and require the other triangle vectors to lie in the $y$-$z$ plane, see figure~\ref{fig:geometry_overview} for a sketch of the relevant vectors. Then we define {\(\mu_1=\cos\theta_1\)} and use $\varphi$, which is the azimuthal angle giving the orientation of the triangle relative to $\n$. \(\theta_{12} = \theta\) is the angle between vectors \(\k_1\) and \(\k_2\), and we define $\mu=\cos\theta=\hat{\k}_1\cdot\hat{\k}_2$.

The bispectrum can then be expressed in terms of five variables, \(\varphi\), \(\mu_1\), \(\theta\), \(k_1\) and \(k_2\), by using
\begin{align}
	\mu_2 &= \sqrt{1-\mu_1^2} \sin\theta \sin\varphi + \mu_1\cos\theta\,,  \\
	\mu_3 &= -\frac{k_1}{k_3}\mu_1 - \frac{k_2}{k_3}\mu_2.
\end{align}
Then 
\begin{equation}\label{eq:bgexpv1}
	B_g(\theta, k_1, k_2, \mu_1, \varphi) = \sum_{\ell m} B_{\ell m}(\theta, k_1, k_2) Y_{\ell m} (\mu_1, \varphi)\,,
\end{equation}
where we use standard orthonormal spherical harmonics,
\begin{equation}
	Y_{\ell m}(\mu_1,\varphi) = \sqrt{\frac{2\ell+1}{4 \pi}} \sqrt{\frac{(\ell-m)!}{(\ell+m)!}} P_\ell^m(\mu_1) e^{{\rm i} m \varphi},
\end{equation}
where the \(P_\ell ^m\) are the associated Legendre polynomials, 
\begin{equation}
	P_\ell ^m(\mu_1) = \frac{(-1)^m}{2^\ell \,\ell !} (1-\mu_1^2)^{m/2} \frac{{\rm d}^{\ell +m}}{{\rm d}\mu_1^{\ell +m}}(\mu_1^2 - 1)^\ell . 
\end{equation}
At this stage we can extract the multipoles numerically once a bias model and cosmological parameters are given. It is actually significantly quicker to perform this extraction algebraically however, as we now explain. 

The bispectrum in general can be considered as a function of
$k_1,k_2,k_3,\mu,\mu_1, \mu_2, \mu_3$ and $\varphi$. An alternative to the expansion~\eqref{eq:bgexpv1} is 
\be\label{eq:bg_absum_reln}
B_g (\mu,k_1,k_2;\mu_1,\mu_2)= \sum_{a=0}^6\sum_{b=0}^6 \mathcal{B}_{ab}(\mu, k_1, k_2)  (\i\mu_1)^a(\i\mu_2)^b\,,
\ee
where we used \(\mu_2\) instead of \(\varphi\) and $a,b=0\ldots6$, which is the maximum power of \(\mu_1,\,\mu_2\) that can arise. This factors out all the angular dependence from the functions $\mathcal{B}_{ab}(\mu,k_1,k_2)$, where $\mu=\cos\theta$, which just depend on the triangle shape (and the cosmology). Note that by explicitly including factors of $\rm i$ in the sum, we have only real coefficients $\mathcal{B}_{ab}$. 
Schematically we can visualise $\mathcal{B}_{ab}$ in matrix form, split into Newtonian and relativistic contributions as (a bullet denotes a non-zero entry, open circles denote zero entries, and dots are non-existent entries; here that means \(a+b > 8\) as higher powers don't occur):
\be \label{eq:bab_newtrelmatrix}
\mathcal{B}_{ab}\sim  \underbrace{ 	
\left( 
\begin {array}{ccccccc}  
\bullet& \circ &\bullet& \circ &\bullet& \circ & \circ \\  
\circ  &\bullet& \circ  &\bullet& \circ &\bullet& \circ \\  
\bullet& \circ &\bullet& \circ &\bullet&  \circ&\bullet\\  
\circ  &\bullet& \circ &\bullet& \circ &\bullet& \cdot \\  
\bullet& \circ &\bullet& \circ &\bullet& \cdot &  \cdot\\  
\circ  &\bullet& \circ &\bullet&  \cdot & \cdot & \cdot \\  
\circ  & \circ &\bullet& \cdot & \cdot & \cdot &  \cdot 
\end {array} 
\right)
}_\text{Newtonian}
+
\underbrace{\left( \begin {array}{ccccccc}   \bullet & \bullet & \bullet & \bullet & \bullet & \bullet & \bullet \\   \bullet & \bullet & \bullet & \bullet & \bullet & \bullet & \bullet \\   \bullet & \bullet & \bullet & \bullet & \bullet & \bullet & \circ  \\   \bullet & \bullet & \bullet & \bullet & \bullet & \circ  & \cdot  \\   \bullet & \bullet & \bullet & \bullet & \circ  & \cdot  & \cdot  \\   \bullet &
 \bullet & \bullet &\circ   & \cdot  & \cdot  & \cdot  \\   \bullet & \bullet & \circ  &\cdot   &  \cdot &  \cdot 
& \cdot  \end {array} \right)}_\text{Relativistic}\,.
\ee
(Note that the matrix row and column labelling start at $a,b=0,0$ for the top left element.) 
Thus, the Newtonian contributions always have $a+b=$\,even\,$\leq8$, contributing only to the real part of \(B_g\), while there are relativistic contributions present for all \(a+b\leq7\). When \(a+b\) is odd, this implies an imaginary component to the full bispectrum.  

In terms of the powers of \(\cH/k\) involved, we can visualise the maximum powers that appear in matrix form as follows:
\be  \label{eq:maxpowerkmatrix}
\mathcal{B}_{ab}\sim \left( \begin {array}{ccccccc} {k}^{-8}&{k}^{-7}&{k}^{-6}&{k}^{-5}&{k
}^{-4}&{k}^{-3}&{k}^{-2}\\  
{k}^{-7}&{k}^{-6}&{k}^{-5
}&{k}^{-4}&{k}^{-3}&{k}^{-2}&{k}^{-1}\\  
{k}^{-6}&{k}
^{-5}&{k}^{-4}&{k}^{-3}&{k}^{-2}&{k}^{-1}&k^0\\  
{k}^{-
5}&{k}^{-4}&{k}^{-3}&{k}^{-2}&{k}^{-1}&k^0 & \cdot\\  
{k}^{-4
}&{k}^{-3}&{k}^{-2}&{k}^{-1}&k^0 &\cdot &\cdot \\  
{k}^{-3}&{k}^{-
2}&{k}^{-1}&k^0 & \cdot&\cdot &\cdot \\  
{k}^{-2}&{k}^{-1}&k^0 &\cdot & \cdot&\cdot & \cdot
\end {array} \right) \,.
\ee

As in the matrix~\eqref{eq:bab_newtrelmatrix}, the matrix row and column labelling in~\eqref{eq:maxpowerkmatrix} starts at \((a,b) = (0,0)\). We see that higher powers \(n\) of \((\cH/k)^n\) appear for lower \(a+b\). Newtonian contributions are all $({\cal H}/k)^0$. Each element has only odd powers of \(\cH/k\) if \(a+b\) is odd, and similarly only even powers if \(a+b\) is even.

The advantage of writing the bispectrum in this form is that we can derive analytic formulas for the multipoles. We need to find
\begin{align}
B_{\ell m} &= \int{\rm d}\Omega \,\,B_g Y^*_{\ell m} \nonumber \\
&= \sum_{a,b} \mathcal{B}_{ab} X_{\ell m}^{ab} \label{eq:xabexp} \,,
\end{align}
where 
\begin{equation} X_{\ell m}^{ab} = \int_0^{2\pi} \diff\varphi\, \int_{-1}^1 \diff\mu_1\, (\i \mu_1)^a (\i \mu_2)^b \,Y_{\ell m}^*(\mu_1,\varphi).
\end{equation}
To do this we use the identity, derived in appendix~\ref{sec:derivationsumappendix}, for \(m \geq 0\),
\begin{align}\label{eq:sumformula} X_{\ell m}^{ab} &= 2^{\ell+m-1}\i^{a+b+m}\sqrt{\frac{\pi(2\ell+1)(\ell-m)!}{(\ell+m)!}}\nonumber\\
&\times
\sum_{p=m}^{\frac{1}{2}(b+m)}\sum_{q=m}^{\ell}
\frac{[1+(-1)^{a+b+q}]\,b!\,\cos^{b+m-2p}\theta\sin^{2p-m}\theta}{4^p(b+m-2p)!(\ell-q)!(p-m)!(q-m)!}
\frac{\Gamma[\frac{1}{2}(q+\ell+1)]}{\Gamma[\frac{1}{2}(q-\ell+1)]}
\frac{\Gamma[\frac{1}{2}(a+b +q-2p+1)]}{\Gamma[\frac{1}{2}(a+b+q+3)]}
\end{align}
for $m\leq b$ and zero otherwise. For $m < 0$, the result follows a similar pattern, using the simple relation \(X_{\ell -m}^{ab} = (-1)^{a+b+m} X_{\ell m}^{ax`'b*}\), see appendix~\ref{sec:derivationsumappendix}. 

The resulting expressions for $B_{\ell m}$ are rather massive, in part because the cyclic permutations become mixed together, so we do not present them here. We can visualise these in matrix form split into their Newtonian and relativistic contributions:
\be \label{eq:blm_newtrelmatrix}
B_{\ell m}=
 \underbrace{\left( \begin {array}{ccccccccc} \bullet& \cdot  &  \cdot& \cdot  &  \cdot& \cdot  &  \cdot& \cdot  &  \cdot
\\  \circ & \circ  &  \cdot& \cdot  &  \cdot& \cdot  &  \cdot& \cdot  &  \cdot \\ \bullet&\bullet&\bullet& \cdot  &  \cdot& \cdot 
& \cdot  &  \cdot& \cdot \\  \circ & \circ  &  \circ& \circ  &  \cdot& \cdot  &  \cdot& \cdot  & \cdot \\ \bullet&\bullet&\bullet
&\bullet&\bullet& \cdot  &  \cdot& \cdot  & \cdot \\ \circ  & \circ  &  \circ& \circ  &  \circ& \circ  &  \cdot& \cdot  & \cdot 
\\ \bullet&\bullet&\bullet&\bullet&\bullet&\bullet&\bullet& \cdot  &  \cdot  \\ \circ  & \circ  &  \circ& \circ  &  \circ&  
\circ & \circ  &  \circ & \cdot \\ \bullet&\bullet&\bullet&\bullet&\bullet&\bullet&\bullet& \circ & \circ \end {array} \right)}_\text{Newtonian}
+
\underbrace{\left( \begin {array}{ccccccccc} \bullet&  \cdot&\cdot  &  \cdot&\cdot  &  \cdot&\cdot  &  \cdot&  \cdot
\\  \bullet&\bullet&  \cdot&\cdot  &  \cdot&\cdot  &  \cdot&\cdot  &  \cdot
\\  \bullet&\bullet&\bullet&  \cdot&\cdot  &  \cdot&\cdot  &  \cdot&  \cdot
\\  \bullet&\bullet&\bullet&\bullet&  \cdot&\cdot  &  \cdot&\cdot  &  \cdot
\\  \bullet&\bullet&\bullet&\bullet&\bullet&  \cdot&\cdot  &  
\cdot & \cdot \\  \bullet&\bullet&\bullet&\bullet&\bullet& \bullet &  \cdot&\cdot  & \cdot \\  \bullet&\bullet&\bullet&\bullet&
\bullet&\bullet&\bullet&  \cdot&  \cdot \\  \bullet&\bullet& \bullet &\bullet&\bullet&\bullet&\bullet& \circ & \cdot \\  \circ  & \circ 
&  \circ&\circ  &  \circ&\circ  &  \circ&\circ  & \circ \end {array} \right)}_\text{Relativistic}\,.
\ee
Again, the matrix indices start at \((0,0)\) in the top left, \((\ell,m) = (0,0)\). In the matrix~\eqref{eq:blm_newtrelmatrix}, consistent with previous matrix visualisations, a closed bullet represents a non-zero entry, while an open circle denotes a vanishing entry. The dots denote the non-existent elements of the matrix, here they are matrix elements where \(m > \ell\) and hence do not exist.
So, the Newtonian bispectrum only induces even multipoles up to and including \(\ell=8\), while the relativistic part induces even and odd multipoles up to $\ell=7$ with multipoles higher than \(\ell = 8\) vanishing exactly. Both the Newtonian and the relativistic part terminate at $m=\pm6$, because $m\leq b\leq6$, as can  be seen from~\eqref{eq:sumformula}. Note that for $m<0$ the pattern is the same.	
In terms of $({\cal H}/k)$ powers, the highest that appear for each $\ell$ is $({\cal H}/k)^{8-\ell}$, while the leading contribution is $({\cal H}/k)^{0~\text{or}~1}$ if the leading contribution is Newtonian or relativistic. These powers are even (odd) if $\ell$ is even (odd), as explained previously along with the visualisation of the powers \(\cH/k\) in equation~\eqref{eq:maxpowerkmatrix}.

\subsection{Presentation of the matrix $\mathcal{B}_{ab}$}

Here we describe in more detail how to calculate the matrix of coefficients $\mathcal{B}_{ab}$. These are far too large to write down, but most of the complexity comes from the $k_i$ permutations and the fact that they are made irreducible from substituting for $\mu_3$. However, the core part can be shown from which they can easily be calculated. First we note that once $\mu_3$ is substituted for, we can write the first cyclic permutation of the product of the kernels as  
\begin{equation}
	\ko_{123}=\ko^{(1)}(k_1,\mu_1)\ko^{(1)}(k_2,\mu_2)\ko^{(2)}(k_1,k_2,k_3,\mu_1,\mu_2)= \sum_{a=0}^5\sum_{b=0}^5 (\i \mu_1)^a (\i \mu_2)^b \ko_{ab}(k_1,k_2,k_3)\,,
\end{equation}
where \(\ko_{ab}(k_1,k_2,k_3)=\ko_{ab}(k_2,k_1,k_3)\) is a set of real \(\mu\)-independent coefficients which we give below, and here the maximum value of \(a,\,b = 5\). Given $\ko_{123}$ we can derive the permutations $\ko_{321}$ and $\ko_{312}$ as
\begin{align}
	&\ko_{321} = \sum_{a,b} \sum_{c=0}^a \binom{a}{c} \frac{k_1^{a-c} k_2^c}{k_3^a} (-1)^a (\i \mu_1)^{a-c} (\i \mu_2)^{b+c} \ko_{ab}(k_3,k_2,k_1)\label{eq:ko_321perm}, \\
	&\ko_{312} = \sum_{a,b} \sum_{c=0}^a \binom{a}{c} \frac{k_1^{a-c} k_2^c}{k_3^a} (-1)^a (\i \mu_1)^{a+b-c} (\i \mu_2)^{c} \ko_{ab}(k_3,k_1,k_2),\label{eq:ko_312perm}
\end{align} 
where, as in general, the range of \(a,\,b = 0 \dots 6\).
Given these, the full bispectrum is just $B_g=\ko_{123}P_1P_2+2\,$permutations, but now explicitly written in terms of sums over powers of $\mu_1,\mu_2$.  From this $\mathcal{B}_{ab}$ can be found by inspection. The difference in dimension between the permutations originates from the other cyclic permutations being added, where one substitutes \(\mu_3 = - \left( k_1 \mu_1 + k_2 \mu_2 \right)/k_3\). In~\eqref{eq:ko_321perm}  the largest power of $\mu_2$ is 6, and~\eqref{eq:ko_312perm} has the largest power of $\mu_1$ as 6.

To present $\ko_{ab}(k_1,k_2,k_3)$ we will show powers of ${\cal H}/k$ separately, and write $\ko_{ab}(k_1,k_2,k_3)=\sum_{n=0}^8 \ko_{ab}^{(n)}(k_1,k_2,k_3)$ where $n$ represents the power of ${\cal H}/k$.  Then the Newtonian and leading GR correction  part look like (again, a bullet denotes a non-zero entry)
\be         
  \ko_{ab}^{(0)} = \left( \begin {array}{cccccc} 
 \bullet&\circ &\bullet&\circ &\bullet&\circ \\ 
  \circ &\bullet&\circ &\bullet &\circ &\bullet\\ 
   \bullet&\circ &\bullet&\circ &\bullet&\circ \\ 
    \circ &\bullet&\circ &\bullet&\circ &\bullet\\ 
 \bullet&\circ &\bullet&\circ &\bullet&\circ \\ 
  \circ &\bullet&\circ &\bullet&\circ &\circ 
\end {array} \right)~~~~~ 
\ko_{ab}^{(1)} = \left( \begin {array}{cccccc} 
  \circ&\bullet&\circ&\bullet&\circ&\bullet\\ \bullet&\circ&\bullet&\circ
&\bullet&\circ\\ \circ&\bullet&\circ&\bullet&\circ&\bullet\\ \bullet&\circ&\bullet&\circ&\bullet&\circ
\\ \circ&\bullet&\circ&\bullet&\circ&\circ\\ \bullet&\circ&\bullet&\circ&\circ&\circ
\end {array}
  \right)
\ee
where, writing $F=F_2(\bm k_1,\bm k_2)$, $G=G_2(\bm k_1,\bm k_2)$, $S=S_2(\bm k_1,\bm k_2)$,  
\begin{align}
\ko^{(0)}_{00}&= 
b_{1}^{2}\left(b_{s^2} S+b_{2}\right)+F b_{1}^{3}
 \\ \ko^{(0)}_{02}&= - b_1f
 \left[b_{1}^{2}+b_{s^2} S +b_{2}+\left(F+\frac{G k_{2}^{2}}{k_{3}^{2}}\right) b_{1}\right]
 \\ \ko^{(0)}_{04}&= 
b_1f^2 \left(\frac{G k_{2}^{2} }{k_{3}^{2}}+b_{1}\right)
 \\ \ko^{(0)}_{11}&= 
-b_1^2f\left[\frac{\left(k_{1}^{2}+k_{2}^{2}\right) b_{1}}{k_{1} k_{2}}+\frac{2 G k_{1}  k_{2} }{k_{3}^{2}}\right] 
 \\ \ko^{(0)}_{13}&= 
 b_1f^2\left[\frac{\left(k_{1}^{2}+2 k_{2}^{2}\right) b_{1}}{k_{1} k_{2}}+\frac{2 G k_{1}  k_{2} }{k_{3}^{2}}\right]
 \\ \ko^{(0)}_{15}&= 
-\frac{b_{1}f^{3} k_{2} }{k_{1}}
 \\ \ko^{(0)}_{20}&= 
- b_1 f \left[b_{1}^{2}+b_{s^2} S + b_{2}+\left(F+\frac{G k_{1}^{2}}{k_{3}^{2}}\right) b_{1}\right]
 \\ \ko^{(0)}_{22}&= 
f^{2}\left[4 b_{1}^{2}+b_{s^2} S+b_{2}+\left(F+\frac{G\left(k_{1}^{2}+k_{2}^{2}\right)}{k_{3}^{2}}\right) b_{1}\right]
 \\ \ko^{(0)}_{24}&= 
 -f^{3}\left(\frac{G k_{2}^{2}}{k_{3}^{2}}+3 b_{1}\right) 
 \\ \ko^{(0)}_{31}&= 
b_1f^{2}\left[\frac{b_{1}\left(2 k_{1}^{2}+k_{2}^{2}\right) }{k_{1} k_{2}}+\frac{2 G k_{1}  k_{2} }{k_{3}^{2}}\right] 
 \\ \ko^{(0)}_{33}&= 
 -f^{3}\left[\frac{2b_{1}\left(k_{1}^{2}+k_{2}^{2}\right) }{k_{1} k_{2}}+\frac{2 G k_{1}  k_{2}}{k_{3}^{2}}\right]
 \\ \ko^{(0)}_{35}&= 
\frac{f^{4} k_{2}}{k_{1}}
 \\ \ko^{(0)}_{40}&= 
b_{1}f^{2}\left(b_1+\frac{G k_{1}^{2} }{k_{3}^{2}}\right) 
 \\ \ko^{(0)}_{42}&= 
 -f^3\left(3b_{{1}}+ {\frac {Gk_{{1}}^{2}}{k_{{3}}^{2}}}
 \right)
 \\ \ko^{(0)}_{44}&= 
2 f^{4}
 \\ \ko^{(0)}_{51}&= 
-\frac{ b_{1}f^{3} k_{1}}{k_{2}}
 \\ \ko^{(0)}_{53}&= 
\frac{f^{4} k_{1}}{k_{2}}.
\end{align}
Similarly, the leading GR correction \(\mathcal{O}(\cH/k)\)  coefficients are, 
\begin{align}
\ko^{(1)}_{01}=&
 b_{1}\gamma_{1}\frac{\left( b_{2}+b_{s^2} S \right)}{k_{2}}
+ b_{1}^{2}
\left(\frac{F \gamma_{1}+\beta_{16}}{k_{2}}+\frac{\beta_{15} \mu}{k_{1}}+\frac{\beta_{14}k_{2} }{k_{1}^{2}}-\frac{  \beta_{19}G k_{2}}{k_{3}^{2}}\right)
 \\ \ko^{(1)}_{03}=& b_{1}f\left[\frac{\left(\beta_{19}-\gamma_{1}\right) G k_{2}}{k_{3}^{2}}-\frac{\beta_{16}}{k_{2}}-\frac{\beta_{15} \mu}{k_{1}}-\frac{\beta_{14} k_{2} }{k_{1}^{2}}\right] 
 -b_{1}^{2}\left(\frac{f \gamma_{1}}{k_{2}}+\frac{ \beta_{17}k_{2}}{k_{1}^{2}}\right) 
 \\ \ko^{(1)}_{05}=&  \frac{b_{1}f \beta_{17}k_{2} }{k_{1}^{2}}
 \\ \ko^{(1)}_{10}=& b_{1}\gamma_{1}\frac{\left(b_{2}+b_{s^2} S \right)}{k_{1}}+  b_{1}^{2}\left[\left(-\frac{G \beta_{19}}{k_{3}^{2}}+\frac{\beta_{14}}{k_{2}^{2}}\right) k_{1}+\frac{\beta_{15} \mu}{k_{2}}+\frac{F \gamma_{1}+\beta_{16}}{k_{1}}\right]
 \\ \ko^{(1)}_{12}=&  - \gamma_{1} f\frac{\left(b_{s^2} S+b_{2}\right)}{k_{1}}-  b_{1}^{2}\left[\gamma_{1} f\left(\frac{k_{1}}{k_{2}^{2}}+\frac{2}{k_{1}}\right) +\frac{\beta_{18}}{k_{1}}\right]\\&
 + b_{1}f\left\{\left[\left(-\frac{2 k_{1}}{k_{3}^{2}}-\frac{k_{2}^{2}}{k_{1} k_{3}^{2}}\right) \gamma_{1}+\frac{k_{1} \beta_{19}}{k_{3}^{2}}\right] G-\frac{F \gamma_{1}}{k_{1}}-\frac{\beta_{14} k_{1}}{k_{2}^{2}}-\frac{\beta_{15} \mu}{k_{2}}-\frac{\beta_{16}}{k_{1}}\right\}
 \\ \ko^{(1)}_{14}=&  b_{1}f\frac{2 \gamma_{1} f+\beta_{18}}{k_{1}}+\frac{G k_{2}^{2} \gamma_{1} f^{2}}{k_{1} k_{3}^{2}}
 \\ \ko^{(1)}_{21}=& -\gamma_{1} f\frac{\left(b_{s^2} S+ b_{2}\right)}{k_{2}}
  -b_{1}^{2}\left[\gamma_1 f\left(\frac{2}{k_{2}}+\frac{k_{2}}{k_{1}^{2}}\right) +\frac{\beta_{18}}{k_{2}}\right] \\
 &+b_1f\left\{\left[\left(-\frac{k_{1}^{2}}{k_{2} k_{3}^{2}}-\frac{2 k_{2}}{k_{3}^{2}}\right) \gamma_{1}+\frac{k_{2} \beta_{19}}{k_{3}^{2}}\right] G-\frac{F \gamma_{1}}{k_{2}}-\frac{\beta_{16}}{k_{2}}-\beta_{14}\left(\frac{ \mu}{k_{1}}+\frac{k_{2} }{k_{1}^{2}}\right)\right\}
 \\ \ko^{(1)}_{23}=&  b_{1}f\left[\left(\frac{4}{k_{2}}+\frac{2 k_{2}}{k_{1}^{2}}\right) \gamma_{1} f+\frac{\beta_{18}}{k_{2}}+\frac{\beta_{17} k_{2} }{k_{1}^{2}} \right]+f^{2}\left[G\left(3 \gamma_{1}-\beta_{19}\right) \frac{k_{2} }{k_{3}^{2}}+\frac{\beta_{16}}{k_{2}}+\frac{\beta_{15} \mu}{k_{1}}+\frac{\beta_{14} k_{2} }{k_{1}^{2}}\right]
 \\ \ko^{(1)}_{25}=& -f^2(\gamma_1f+\beta_{17})\frac{k_{2} }{k_{1}^{2}}
 \\ \ko^{(1)}_{30}=& - b_{1}^{2}\left(\frac{f \gamma_{1}}{k_{1}}+\frac{\beta_{17} k_{1} }{k_{2}^{2}}\right)+b_{1}f\left[G\left(- \gamma_{1}+\beta_{19}\right)  \frac{k_{1} }{k_{3}^{2}}-\frac{\beta_{14} k_{1}}{k_{2}^{2}}-\frac{\beta_{15} \mu}{k_{2}}-\frac{\beta_{16}}{k_{1}}\right] 
 \\ \ko^{(1)}_{32}=& b_{1}f\left[\left(\frac{2 k_{1}}{k_{2}^{2}}+\frac{4}{k_{1}}\right) \gamma_{1} f+\frac{ \beta_{17} k_{1}}{k_{2}^{2}}+\frac{\beta_{18}}{k_{1}}\right] +f^2\left[G\left(3 \gamma_{1}- \beta_{19}\right) \frac{k_{1} }{k_{3}^{2}}+\frac{\beta_{14} k_{1}}{k_{2}^{2}}+\frac{\beta_{15} \mu}{k_{2}}+\frac{\beta_{16}}{k_{1}}\right]
 \\ \ko^{(1)}_{34}=& -\frac{f^{2}\left(3 f \gamma_{1}+\beta_{18}\right)}{k_{1}}
 \\ \ko^{(1)}_{41}=& b_{1}f\frac{\left(2 \gamma_{1} f+ \beta_{18}\right)}{k_{2}} +\frac{G \gamma_{1} f^{2} k_{1}^{2} }{k_{2} k_{3}^{2}}
 \\ \ko^{(1)}_{43}=&-\frac{f^{2}\left(3 f \gamma_{1}+\beta_{18}\right)}{k_{2}}
 \\ \ko^{(1)}_{50}=& \frac{b_1 \beta_{17} f k_1}{k_2^2}
 \\ \ko^{(1)}_{52}=&-\frac{f^{2} k_{1}\left(f \gamma_{1}+\beta_{17}\right)}{k_{2}^{2}}\,.
\end{align}
The remaining matrices are of the form 
\begin{align}
 \ko^{(2)}_{ab}&= 
 \left( \begin {array}{cccccc} \bullet &\circ &\bullet &\circ 
&\bullet &\circ \\  \circ &\bullet &\circ &
\bullet &\circ &\bullet \\  \bullet &\circ 
&\bullet &\circ &\bullet &\circ \\  \circ &
\bullet &\circ &\bullet &\circ &\circ \\  
\bullet &\circ &\bullet &\circ &\circ &\circ 
\\  \circ &\bullet &\circ &\circ &\circ &
\circ \end {array} \right) 
~~~ \ko^{(3)}_{ab}= 
 \left( \begin {array}{cccccc} \circ &\bullet &\circ &\bullet 
&\circ &\bullet \\  \bullet &\circ &
\bullet &\circ &\bullet &\circ \\  \circ &
\bullet &\circ &\bullet &\circ &\circ \\  
\bullet &\circ &\bullet &\circ &\circ &\circ 
\\  \circ &\bullet &\circ &\circ &\circ &
\circ \\  \bullet &\circ &\circ &\circ &
\circ &\circ \end {array} \right) 
~~~ \ko^{(4)}_{ab}= 
 \left( \begin {array}{cccccc} \bullet &\circ &\bullet &\circ 
&\bullet &\circ \\  \circ &\bullet &\circ &
\bullet &\circ &\circ \\  \bullet &\circ &
\bullet &\circ &\circ &\circ \\  \circ &
\bullet &\circ &\circ &\circ &\circ \\  
\bullet &\circ &\circ &\circ &\circ &\circ 
\\  \circ &\circ &\circ &\circ &\circ &
\circ \end {array} \right) 
~~~ \ko^{(5)}_{ab}= 
 \left( \begin {array}{cccccc} \circ &\bullet &\circ &\bullet 
&\circ &\circ \\  \bullet &\circ &\bullet &
\circ &\circ &\circ \\  \circ &\bullet &
\circ &\circ &\circ &\circ \\  \bullet &
\circ &\circ &\circ &\circ &\circ \\  
\circ &\circ &\circ &\circ &\circ &\circ 
\\  \circ &\circ &\circ &\circ &\circ &
\circ \end {array} \right) \nonumber
\\ \ko^{(6)}_{ab}&= 
 \left( \begin {array}{cccccc} \bullet &\circ &\bullet &\circ 
&\circ &\circ \\  \circ &\bullet &\circ &
\circ &\circ &\circ \\  \bullet &\circ &
\circ &\circ &\circ &\circ \\  \circ &
\circ &\circ &\circ &\circ &\circ \\  
\circ &\circ &\circ &\circ &\circ &\circ 
\\  \circ &\circ &\circ &\circ &\circ &
\circ \end {array} \right) 
~~~ \ko^{(7)}_{ab}= 
 \left( \begin {array}{cccccc} \circ &\bullet &\circ &\circ &
\circ &\circ \\  \bullet &\circ &\circ &
\circ &\circ &\circ \\  \circ &\circ &
\circ &\circ &\circ &\circ \\  \circ &
\circ &\circ &\circ &\circ &\circ \\  
\circ &\circ &\circ &\circ &\circ &\circ 
\\  \circ &\circ &\circ &\circ &\circ &
\circ \end {array} \right) 
~~~ \ko^{(8)}_{ab}= 
 \left( \begin {array}{cccccc} \bullet &\circ &\circ &\circ &
\circ &\circ \\  \circ &\circ &\circ &
\circ &\circ &\circ \\  \circ &\circ &
\circ &\circ &\circ &\circ \\  \circ &
\circ &\circ &\circ &\circ &\circ \\  
\circ &\circ &\circ &\circ &\circ &\circ 
\\  \circ &\circ &\circ &\circ &\circ &
\circ \end {array} \right).
\end{align}
Their coefficients are extracted in similar fashion, and can be found in full in appendix~\ref{sec:kabappendix}.

\section{Analysis}\label{sec:anal}

Here we present an analysis of the behaviour of the  multipoles.

\subsection{ Co-linear, squeezed and equilateral limits}

\begin{figure}[H]
\centering
\includegraphics[width=0.5\textwidth]{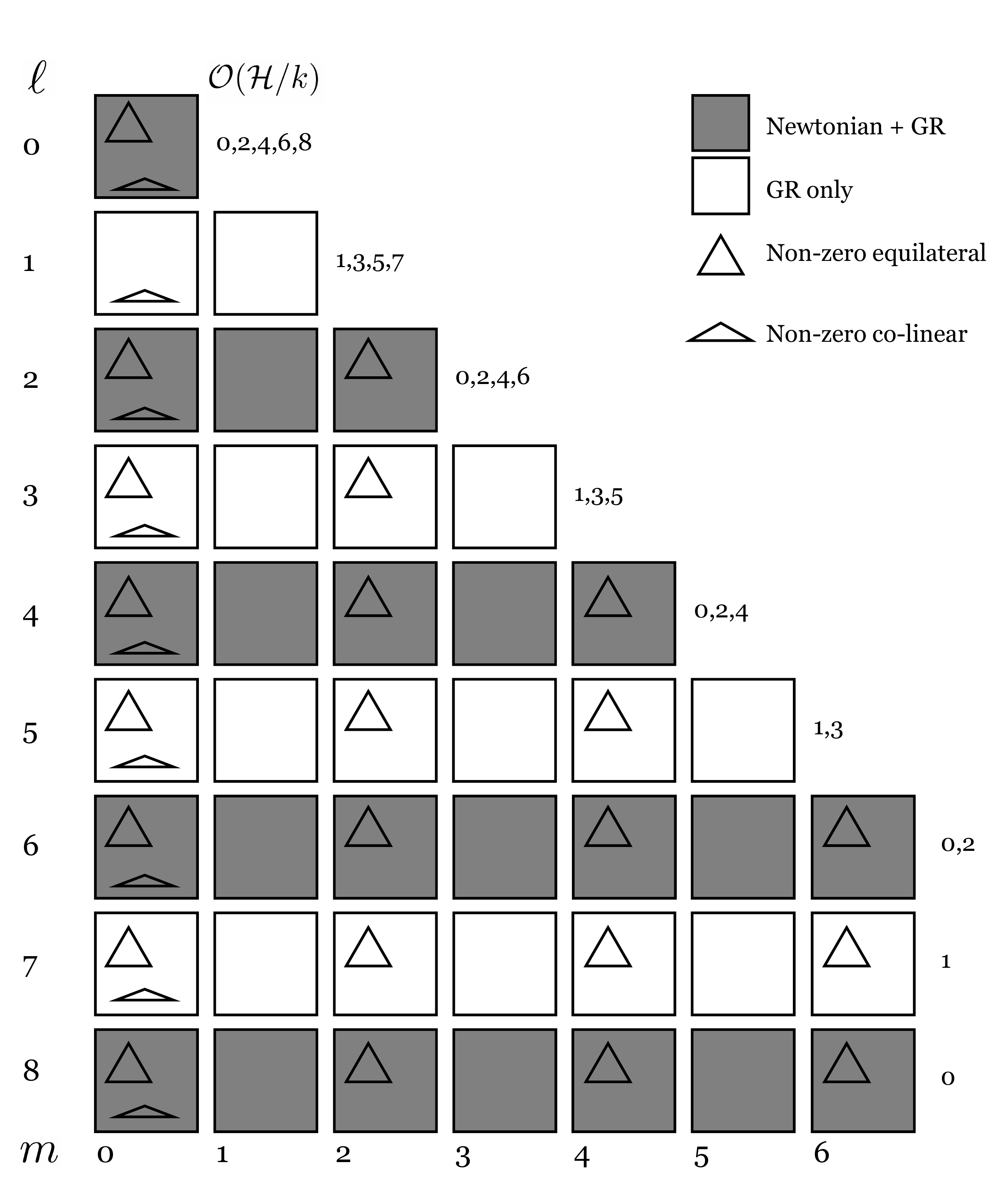}
\caption{Overview of all non-zero multipoles for the bispectrum, which includes \(\ell\) from 0 to 8, and \(m\) from \(-\ell\) to \(\ell\); the pattern here is the same for \(m < 0\), so only $m \geq 0$ are displayed. Denoted in the figure are whether components are Newtonian+GR or GR only, triangle shapes indicating whether given components are non-vanishing in flattened (co-linear) or equilateral limits. Note how the dipole is unique in having the equilateral case vanish for every value of \(m\). Also given is which powers of \(\cH/k\) appear in each of the multipoles. \label{fig:blm_overview}}
\end{figure}

To help understand further the multipoles we can evaluate their equilateral $(k_1=k_2=k_3)$, co-linear   ($\theta=0$ or $\theta=\pi$) and squeezed limits analytically.  
Non-zero co-linear multipoles exist only for \(m=0\) components. This is the one limit that is easy to evaluate by hand~-- it follows directly from~\eqref{eq:sumformula}. 
The equilateral case is significantly more complicated to evaluate.
Non-zero equilateral multipoles exist for all even \(m\), for any $\ell$, the one exception being the \(m=0\) part of the dipole, for which the equilateral configuration is identically zero. These are summarised in Fig.~\ref{fig:blm_overview}, together with the powers of $k$ which appear in each multipole.

The squeezed limit was explicitly evaluated in~\cite{Clarkson:2018dwn} for the leading $\mathcal{O}(\mathcal{H}/k_L)$ contribution, where $k_L$ is the long mode, which we expand further here. {Note that in what follows, we have assumed that the small-scale modes are sufficiently sub-equality scale, and that the large-scale modes are larger than the equality scale. The leading corrections in the even multipoles require us going beyond leading order $\mathcal{O}(\mathcal{H}/k_L)$ in the squeezed limit. We let 
\begin{equation}
k_{1}=k_{2}=k_S, ~~k_{3}=\epsilon k_S\,,
\end{equation}
to write the wavenumber in terms of the short mode $k_S\gg k_L$, which implies
\begin{equation}
\mu=-1+\frac{\epsilon^2}{2}\,.
\end{equation}
We then take the limit as $\epsilon\to0$ with the short mode $k_S$ fixed, and keep only the leading terms in $\cH/k_L$, neglecting factors of $\cH/k_S$ and $P(k_S)^2$.  For each multipole we are then left with the squeezed limit as a polynomial in $\cH/k_L$. The leading contributions are:
\be
\underbrace{\left(\frac{\cH}{k_L}\right)^0 
 \left( \begin {array}{ccccccccc} \bullet &\cdot &\cdot &\cdot &\cdot &\cdot &\cdot &\cdot &\cdot 
\\  \circ &\circ &\cdot &\cdot &\cdot &\cdot &\cdot &\cdot &\cdot \\  \bullet &\circ &\bullet &\cdot &\cdot &\cdot 
&\cdot &\cdot &\cdot \\  \circ &\circ &\circ &\circ &\cdot &\cdot &\cdot &\cdot &\cdot \\  \bullet &\circ &\bullet 
&\circ &\bullet &\cdot &\cdot &\cdot &\cdot \\  \circ &\circ &\circ &\circ &\circ &\circ &\cdot &\cdot &\cdot 
\\  \bullet &\circ &\bullet &\circ &\bullet &\circ &\circ &\cdot &\cdot \\  \circ &\circ &\circ &\circ &\circ &\circ 
&\circ &\circ &\cdot \\  \bullet &\circ &\bullet &\circ &\bullet &\circ &\circ &\circ &\circ \end {array} \right) 
}_{\text{Newtonian part}}
+
\underbrace{
 \left(\frac{\cH}{k_L}\right)^1\left( \begin {array}{ccccccccc} {\circ }&\cdot &\cdot &\cdot &\cdot &\cdot &\cdot &\cdot &\cdot \\ \circ &\bullet &\cdot &\cdot &\cdot &\cdot &\cdot &\cdot &\cdot \\ {\circ }&\circ &{\circ }&\cdot &\cdot &\cdot &\cdot &\cdot &\cdot \\ \circ &\bullet &\circ &\bullet &\cdot &\cdot &\cdot &\cdot &\cdot \\ {\circ }&\circ &{\circ }&\circ &\circ &\cdot &\cdot &\cdot &\cdot \\ \circ &\bullet &\circ &\bullet &\circ &\circ &\cdot &\cdot &\cdot \\ {\circ }&\circ &{\circ }&\circ &\circ &\circ &\circ &\cdot &\cdot \\ \circ &\bullet &\circ &\bullet &\circ &\circ &\circ &\circ &\cdot \\ \circ &\circ &\circ &\circ &\circ &\circ &\circ &\circ &\circ \end {array} \right) 
+
\left(\frac{\cH}{k_L}\right)^2 \left( \begin {array}{ccccccccc} \bullet &\cdot &\cdot &\cdot &\cdot &\cdot &\cdot &\cdot &\cdot \\ \circ &\circ &\cdot &\cdot &\cdot &\cdot &\cdot &\cdot &\cdot \\ \bullet &\circ &\bullet &\cdot &\cdot &\cdot &\cdot &\cdot &\cdot \\ \circ &\circ &\circ &\circ &\cdot &\cdot &\cdot &\cdot &\cdot \\ \bullet &\circ &\bullet &\circ &\circ &\cdot &\cdot &\cdot &\cdot \\ \circ &\circ &\circ &\circ &\circ &\circ &\cdot &\cdot &\cdot \\ \bullet &\circ &\bullet &\circ &\circ &\circ &\circ &\cdot &\cdot \\ \circ &\circ &\circ &\circ &\circ &\circ &\circ &\circ &\cdot \\ \circ &\circ &\circ &\circ &\circ &\circ &\circ &\circ &\circ \end {array} \right) }_\text{GR contributions}
\ee
Here, the matrices represent the $\ell,m$ values from $\ell=0,m=0$ (top left entry). We see that the Newtonian part has non-zero squeezed limits for some even $m$, terminating at $m=4$. GR corrections come in up to $m=3$ for $\ell\leq 7$. For odd $m$ these contributions come in for the leading terms $\mathcal{O}(\mathcal{H}/k)$, while for $m$ even the order is lower, $\mathcal{O}((\mathcal{H}/k)^2)$. Note that we assume primordial Gaussianity. In the presence of primordial non-Gaussianity, the squeezed limit has higher powers of \(\cH/k\). Current work investigates how primordial non-Gaussianity will change our results. The effect of local primordial non-Gaussianity on the Newtonian galaxy bispectrum is presented in~\cite{Umeh:2016nuh}.

\subsection{Numerical results}

Here we present a numerical analysis of the multipoles of the galaxy bispectrum. We use three different survey models, two of which are appropriate for future surveys; i.e. SKA HI intensity mapping, and a Stage IV \(H\alpha\) spectroscopic galaxy survey similar to Euclid. The third model we consider is a simplified `toy model' for illustrative purposes. The parameters we use are introduced below. 

Evolution and magnification bias are defined as~\cite{Alonso_2015}, 
\begin{equation}\label{eq:beQdefs}
	b_e = - \frac{\partial \ln n_g}{\partial \ln(1+z)}, \quad \Q = - \evaluat*{\frac{\partial\ln n_g}{\partial \ln L}}_{c},
\end{equation}
where \(n_g\) is the comoving galaxy number density, \(L\) the luminosity, and \(|_c\) denotes evaluation at the flux cut. 

For an HI intensity mapping survey, we estimate the bias from the halo model following~\cite{Umeh:2015gza}.  
This yields the following fitting formulae for first and second order bias, 
\begin{align}
	b^{\mathrm{HI}}_1(z) &=    0.754 + 0.0877 z + 0.0607 z^2 - 0.00274 z^3 \,, \\ 
	b^{\mathrm{HI}}_2(z) &=  -0.308 - 0.0724 z - 0.0534 z^2 + 0.0247 z^3  \,.
\end{align}
For the tidal bias, we assume zero initial tidal bias which relates \(b_{s^2}\) to \(b_1\) as,
\begin{equation}
	b_{s^2}^{\mathrm{HI}}(z) = - \frac{4}{7} (b_1(z) -1),
\end{equation}
so that, 
\begin{equation}
	b_{s^2}^{\mathrm{HI}}(z) = 0.141 - 0.0501 z - 0.0347 z^2 + 0.00157 z^3.
\end{equation}
The HI intensity mapping evolution bias is given by the background HI brightness temperature~\cite{Fonseca:2018hsu},
\begin{equation}
	b_e^{\mathrm{HI}}(z) = - \frac{\diff \ln \left[ (1+z)^{-1} \cH \mean{T}_{\mathrm{HI}}\right]}{\diff \ln\left[1+z \right]}\,,
\end{equation}
where \(\mean{T}_{\mathrm{HI}}\) is given by the fitting formula, 
\begin{equation}
	\mean{T}_{\mathrm{HI}}(z) = (5.5919 + 23.242  z - 2.4136  z^2)\times 10^{-2} \, \mathrm{mK}.
\end{equation}
The effective magnification bias for HI intensity mapping is~\cite{Fonseca:2018hsu}
\begin{equation}
	\Q^{\mathrm{HI}} = 1.0\,,
\end{equation}
and clustering bias is independent of luminosity, 
\begin{equation}
	\frac{\partial b_1^\mathrm{HI}}{\partial \ln L} = 0\,.
\end{equation}

We consider a Stage IV \(H\alpha\) spectroscopic survey similar to Euclid, and use the clustering biases given in~\cite{Maartens:2019yhx}, 
\begin{align}
	b^{H\alpha}_1(z) &= 0.9 + 0.4 z,\\
	b^{H\alpha}_2(z) &= -0.741 - 0.125 z + 0.123 z^2 + 0.00637 z^3,\\
	b^{H\alpha}_{s^2}(z) &= 0.0409 - 0.199 z - 0.0166 z^2 + 0.00268 z^3.
\end{align}
The magnification bias and evolution bias are~\cite{Maartens:2019yhx},
\begin{align}
	\Q^{H\alpha}(z) &= \frac{ y_c(z)^{\alpha + 1} \exp\left[-y_c(z)\right] }{\Gamma(\alpha+1, y_c(z))},\\
	b_e^{H\alpha}(z) &= - \frac{\diff \ln \Phi_\ast (z)}{\diff \ln(1+z)} + \frac{\diff \ln y_c(z)}{\diff \ln (1+z)}\Q^{H\alpha}(z)\,,
\end{align}
where \(\alpha = - 1.35\), \(\Gamma\) is the upper incomplete gamma function, \(\Phi_\ast\) is given in~\cite{Maartens:2019yhx} and \(y_c(z) = \left[ \chi(z) / \left( 2.97 h \times 10^3 \right) (\mathrm{Mpc}/h) \right]^2\). Table 1 in~\cite{Maartens:2019yhx} summarises the numerical values of the bias parameters discussed above. Finally, we follow~\cite{Maartens:2019yhx} and take
\begin{equation}
	\evaluat*{\frac{\partial b_1^{H\alpha}}{\partial \ln L}}_\text{c} = 0\,.
\end{equation}

For the simple model of galaxy bias, we use
\begin{align}
	b_1(z) &= \sqrt{1+z}, \\
	b_2(z) &= -0.3 \sqrt{1+z}, \\
	b_{s^2}(z) &= - \frac{4}{7} (b_1(z) -1),\\
	b_e &= 0, \\
	\Q &= 0.
\end{align}

For cosmological parameters we use Planck 2018~\cite{Aghanim:2018eyx}, giving the best-fit parameters \(h = 0.6766,\,\Omega_{m0}=0.3111,\,\Omega_{b0}h^2=0.02242,\,\Omega_{c0}h^2 = 0.11933,\,n_s = 0.9665,\,\gamma = \ln f / \ln \Omega_m = 0.545\). The linear matter power spectrum is calculated using CAMB~\cite{Lewis:1999bs}.

We examine numerically three different triangular configurations, the squeezed, co-linear, and equilateral triangles, as a function of triangle size. For our numerical analysis, we choose a moderately squeezed triangle shape with \(\theta\approx 178^\circ\), which corresponds to \(k_3 = k, \, k_1 = k_2 = 28k\) (such that long mode \(k_3\) is the reference wavevector, and the other vectors are defined in relation to the long mode). For the co-linear case, we use flattened isosceles triangles with \(\theta  \approx 2.3^\circ\), corresponding to \(k_3 = k, \, k_1 = k_2 = 0.5001 k.\) All plots are at redshift \(z=1\), with the exception of figure~\ref{fig:fnofz}, where we look at the amplitude as a function of redshift. 

Firstly, we consider the total amplitude of the different multipoles with respect to the Newtonian monopole, plotting the total power contained in each of the multipoles and normalising by the Newtonian monopole of the galaxy bispectrum, 
\begin{equation}\label{eq:totpower}
b_\ell(k_1,k_2,\theta) = \frac{1}{ B_{\mathrm{N},00}(k_1,k_2,\theta)}{\sqrt{\frac{1}{2 \ell + 1} \displaystyle\sum_{m=-\ell}^\ell |B_{\ell m}(k_1,k_2,\theta)|^2}}. 
\end{equation}
We present this for all multipoles \(\ell = 0 \ldots 8\) and separately for each of the triangle shapes introduced above (i.e. fixing triangle shape, and varying size by varying \(k\)), as well as for both bias models which are relevant for future surveys. The results can be viewed in figures~\ref{fig:totpowersq},~\ref{fig:totpowerfl} and~\ref{fig:totpowereq}.

We have created colour-intensity plots to give an overview of the relative amplitudes of the first few multipoles of the galaxy bispectrum, \(\ell = 0 \ldots 3\). Because of the simple relationship between \(B_{\ell,m}\) and \(B_{\ell, -m}\), we do not show plots for negative \(m\). These as well are done for both HI intensity mapping bias and \(H\alpha\) bias. The results are shown in figures~\ref{fig:euclidtriangle} and~\ref{fig:imtriangle} for the Euclid-like survey and for SKA intensity mapping respectively. 

To further investigate the dependence on triangle shape we investigate the reduced bispectrum. We define the reduced bispectrum as
\begin{equation}
	Q_{\ell m}(k_1,k_2,\theta) = \frac{B_{\ell m}(k_1,k_2,\theta)}{P_0(k_1) P_0(k_2) + P_0(k_2) P_0(k_3) + P_0(k_1) P_0(k_3)},
\end{equation}
where 
\(P_0\) is the monopole of the galaxy power spectrum,
\begin{equation}
	P_0(k) = \frac{1}{2} \int_{-1}^1 \diff\mu \, P_g(\k), 
\end{equation}
with the galaxy power spectrum \(P_g(\k) = (b_1 + f \mu^2)^2 P\), \(P\) being the linear dark matter power spectrum. (An alternative definition would be to use the relativistic galaxy power spectrum which would induce small changes $O((\cH/k)^2)$ on Hubble scales.) The reduced bispectrum \(Q\) is hence dependent on magnitude of wavevectors \(\ka\) and \(\kb\), and the angle  between these (\(\pi-\theta\)). We fix \(k_1 = \SI{0.1}{\mega\parsec\tothe{-1}}\) and \(k_1 = \SI{0.01}{\mega\parsec\tothe{-1}}\), and use differently coloured lines to indicate the ratio of \(k_2/k_1\), which ranges from isosceles triangles in which \(k_1 = k_2\), to \(k_2/k_1 = 4.5\). The angle \(\theta\) ranges from \([0,\pi]\), except for the isosceles shape, for which we stop at \(\theta = \pi - 0.01\) (for \(k_1 = \SI{0.1}{\mega\parsec\tothe{-1}}\)), and at \(\theta = \pi - 0.02\) (for \(k_1 = \SI{0.01}{\mega\parsec\tothe{-1}}\)). The reason for this is the inclusion of relativistic \(\cH/k\) contributions, which cause unobservable divergences as \(k \to 0\), occurring here for the isosceles shape when the angle between \(\ka\) and \(\kb\) goes to \(\pi\) and \(k_3 \to 0\). 

The bias used is again that for the Euclid-like \(H\alpha\) spectroscopic survey. Results are in figure~\ref{fig:Q_rb}. The layout is similar to figures~\ref{fig:euclidtriangle} and~\ref{fig:imtriangle}, with \(\ell = 0 \ldots 3\) plotted. Once again, negative \(m\) are not shown. 

Lastly, we fix triangle shape and size, and plot the relative total power (as defined in ~\eqref{eq:totpower}) as a function of redshift, where redshift ranges from \( z = 0.1 \ldots 2.0\). This is done for the toy model for bias only. The three panels in figure~\ref{fig:fnofz} show the results for \(\ell = 0 \ldots 3\), for each of the three wavevector triangles discussed earlier; equilateral, squeezed and flattened shapes. Solid and dashed lines indicate the relative total power for \(k_1 = \SI{0.1}{\mega\parsec\tothe{-1}}\) and \(k_1 = \SI{0.01}{\mega\parsec\tothe{-1}}\) respectively. 

\clearpage
\begin{figure}[H]
\begin{subfigure}{0.5\textwidth}
\includegraphics[width=0.85\linewidth]{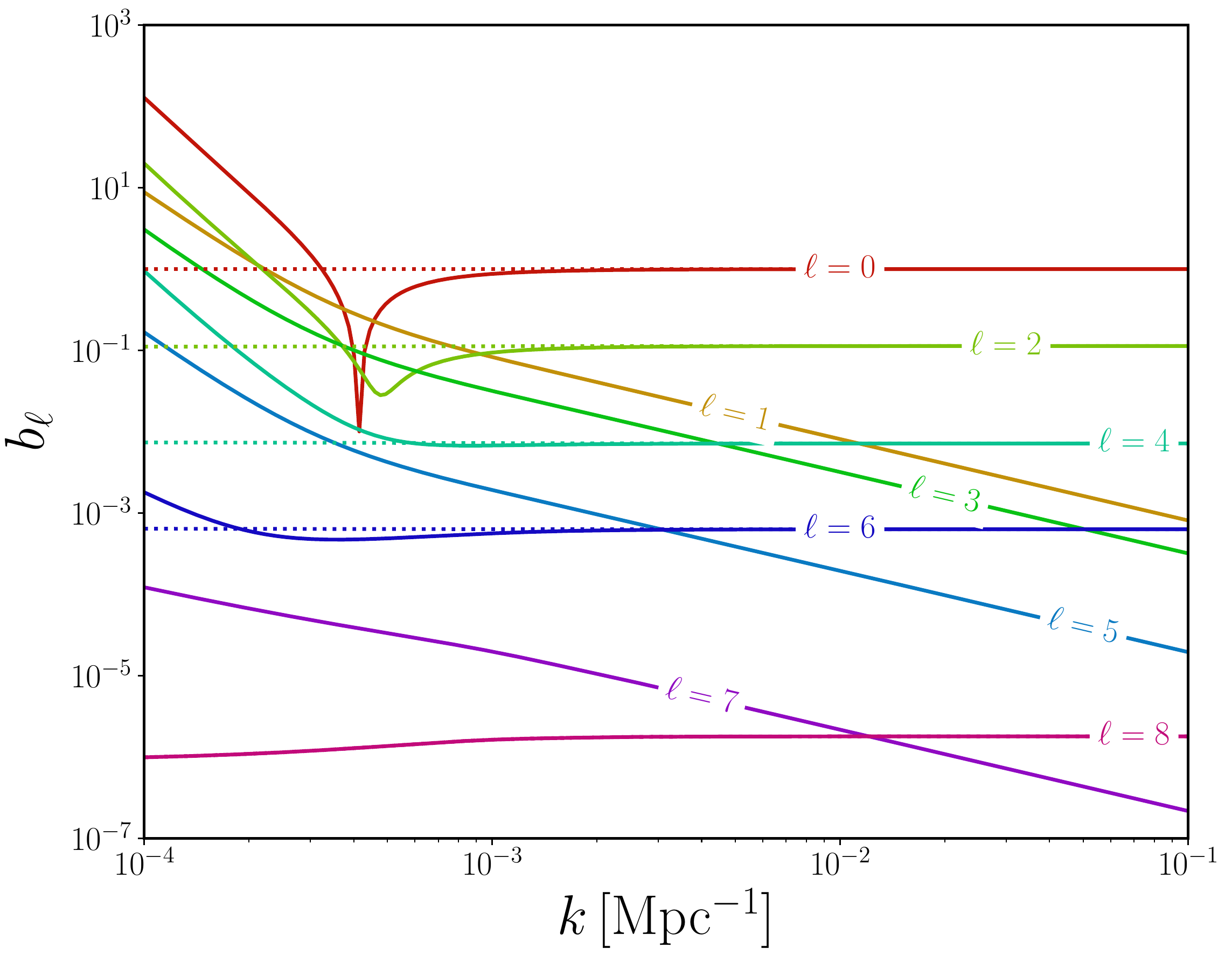}
\end{subfigure}%
\begin{subfigure}{0.5\textwidth}
\includegraphics[width=0.85\linewidth]{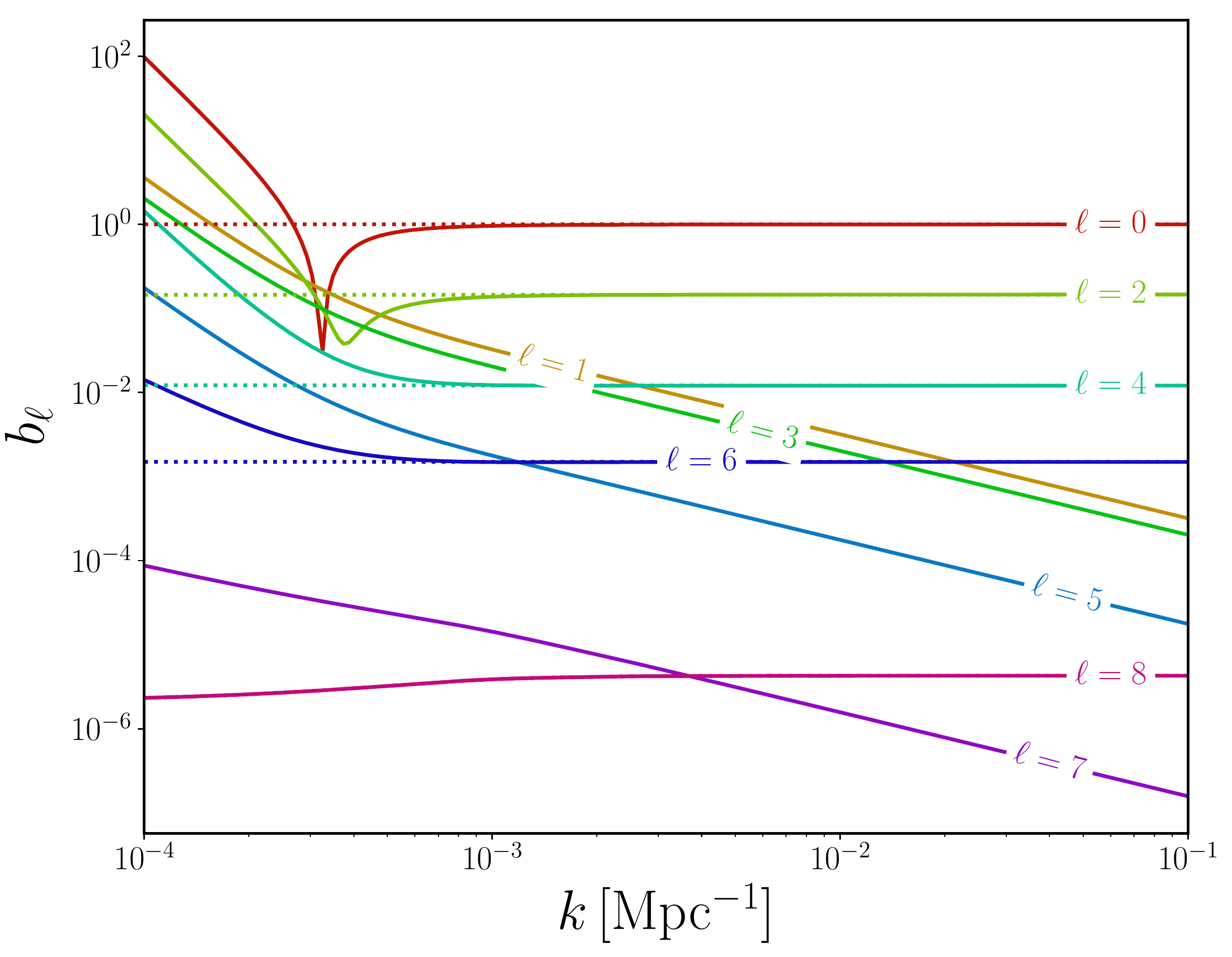}
\end{subfigure}
\caption{Normalised total power for squeezed configuration for Euclid-like (left) and SKA HI intensity mapping (right) surveys. Long mode \(k_3\) is plotted along the \(x\)-axis. \label{fig:totpowersq}}
\end{figure}
\begin{figure}[H]
\begin{subfigure}{0.5\textwidth}
\includegraphics[width=0.85\linewidth]{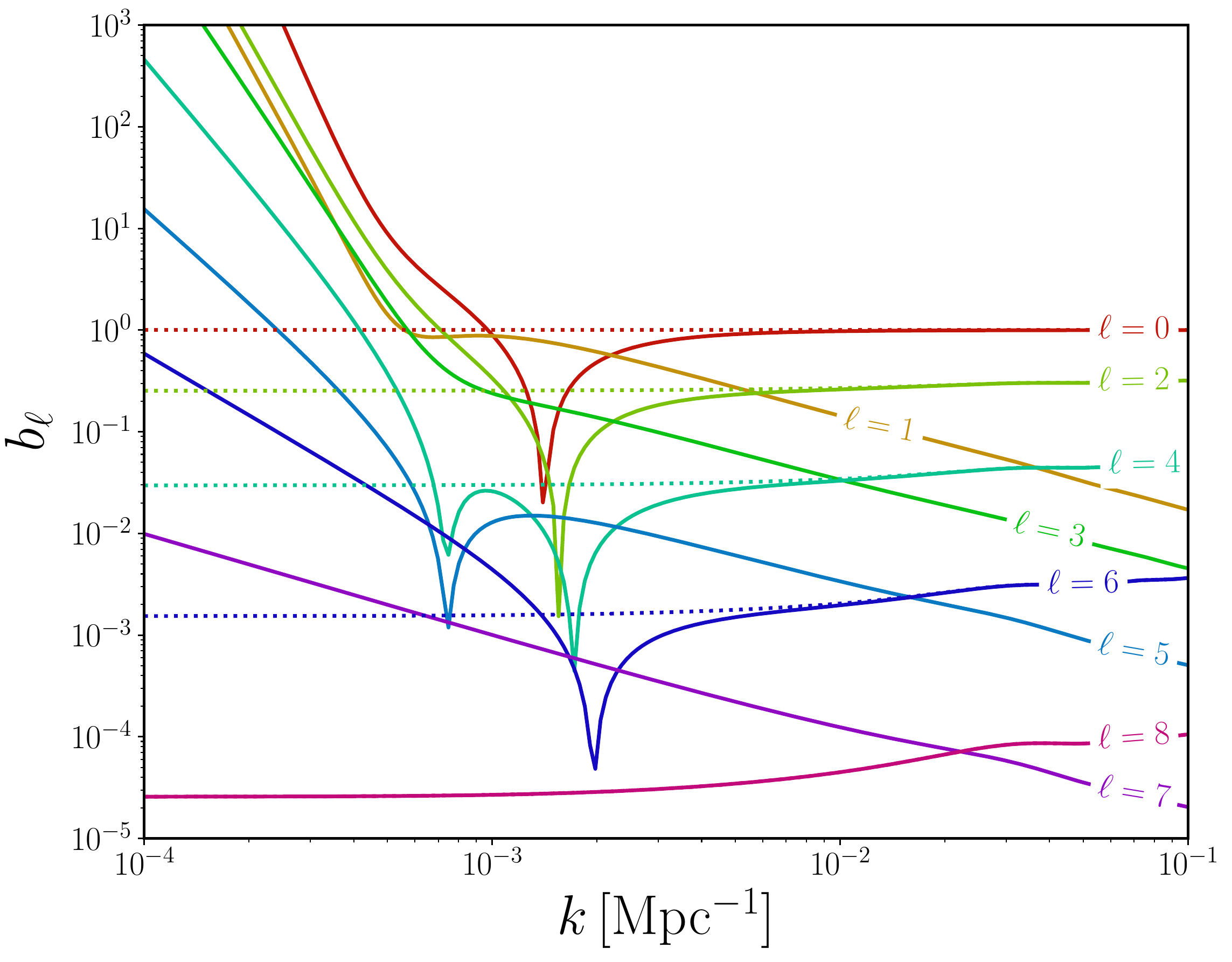}
\end{subfigure}%
\begin{subfigure}{0.5\textwidth}
\includegraphics[width=0.85\linewidth]{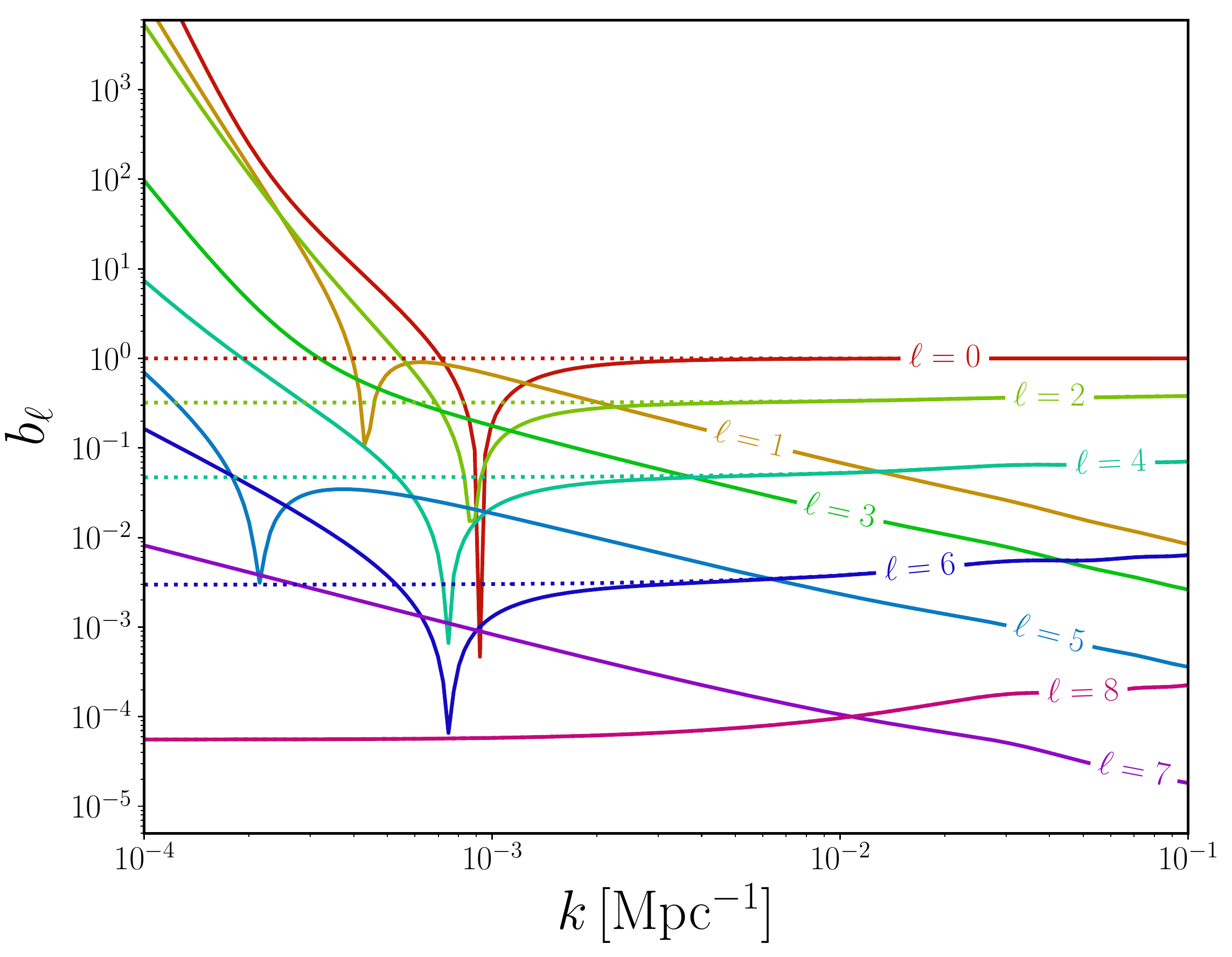}
\end{subfigure}
\caption{Total power for flattened configuration for Euclid-like (left) and SKA HI intensity mapping (right) surveys. Long mode \(k_3\) is plotted along the \(x\)-axis. \label{fig:totpowerfl}}
\end{figure}
\begin{figure}[H]
\begin{subfigure}{0.5\textwidth}
\includegraphics[width=0.85\linewidth]{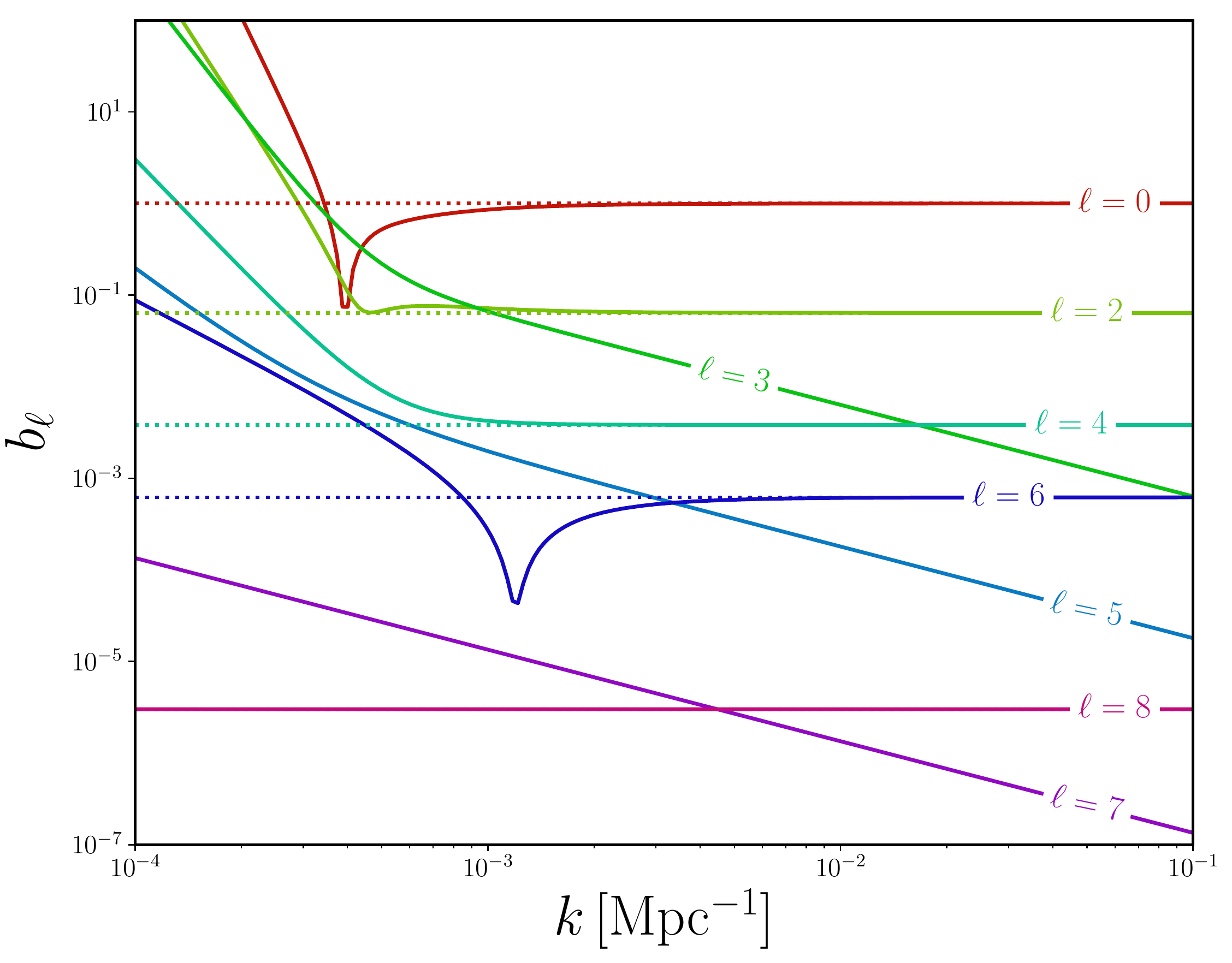}
\end{subfigure}%
\begin{subfigure}{0.5\textwidth}
\includegraphics[width=0.85\linewidth]{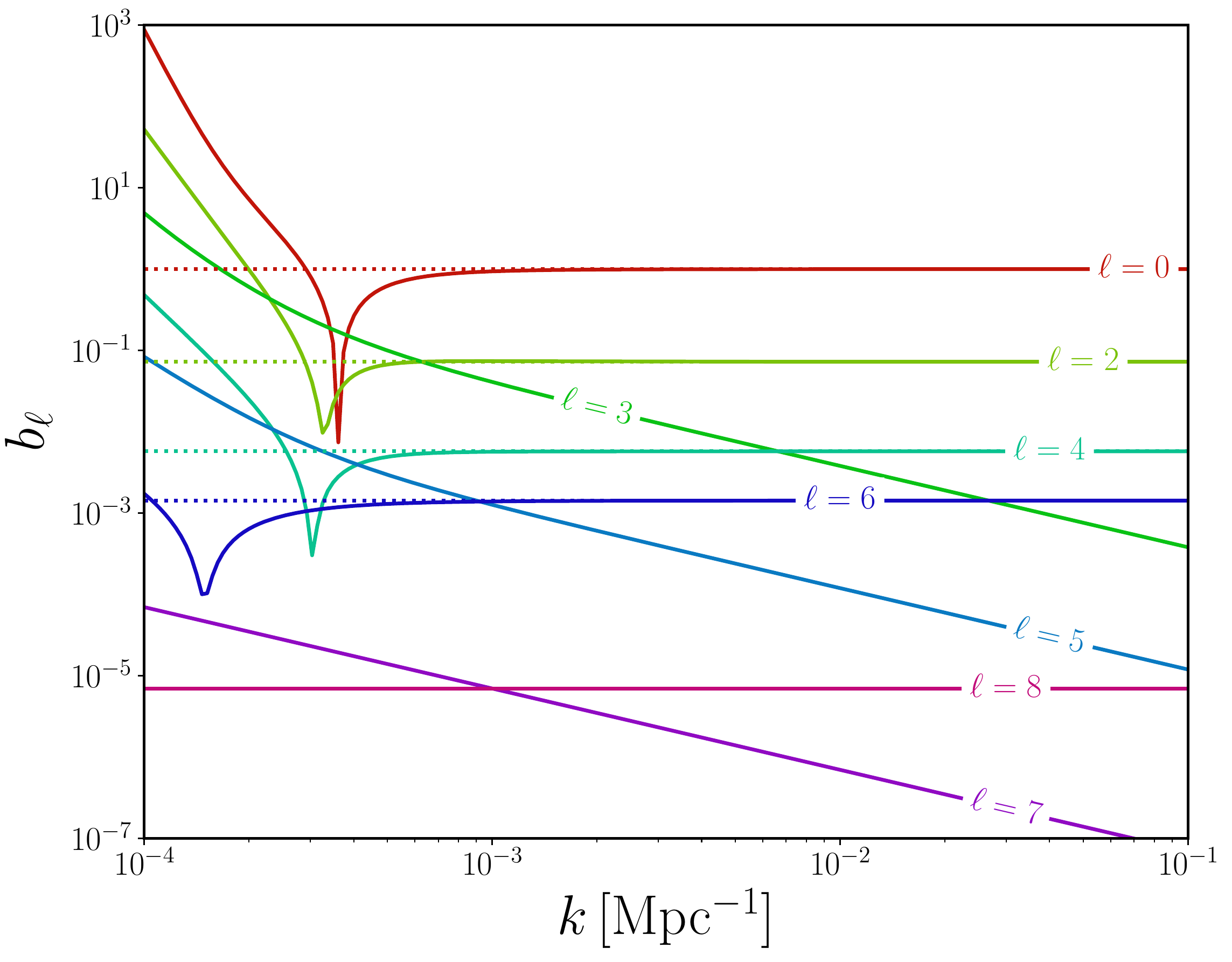}
\end{subfigure}
\caption{Total power for equilateral configuration for Euclid-like (left) and SKA HI intensity mapping (right) surveys. Since the dipole vanishes in this limit, the \(\ell = 1\) line is absent.\label{fig:totpowereq}}
\end{figure}

\clearpage
\begin{figure}[H]
\centering
    \includegraphics[width=0.8\textwidth]{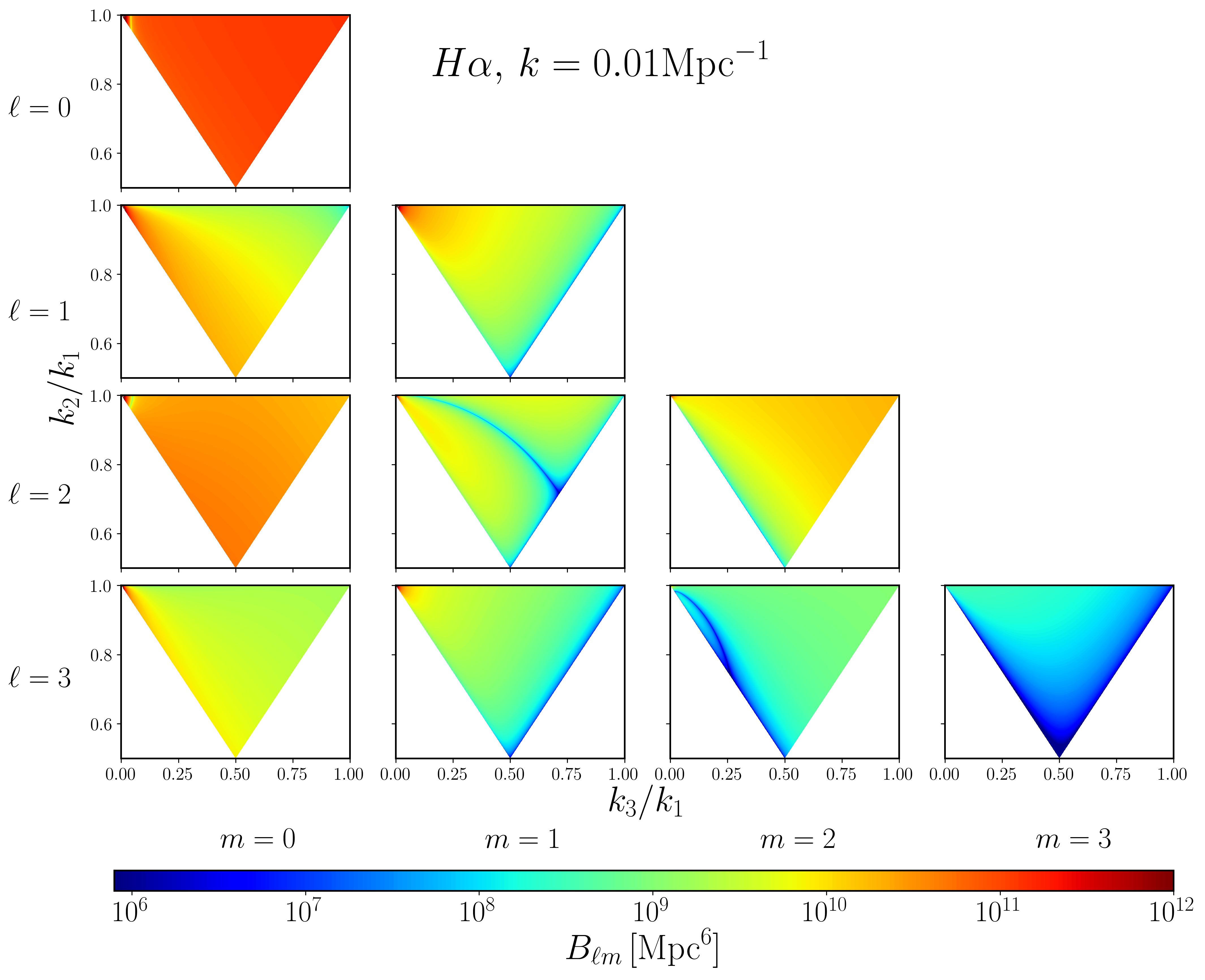}
	\includegraphics[width=0.8\textwidth]{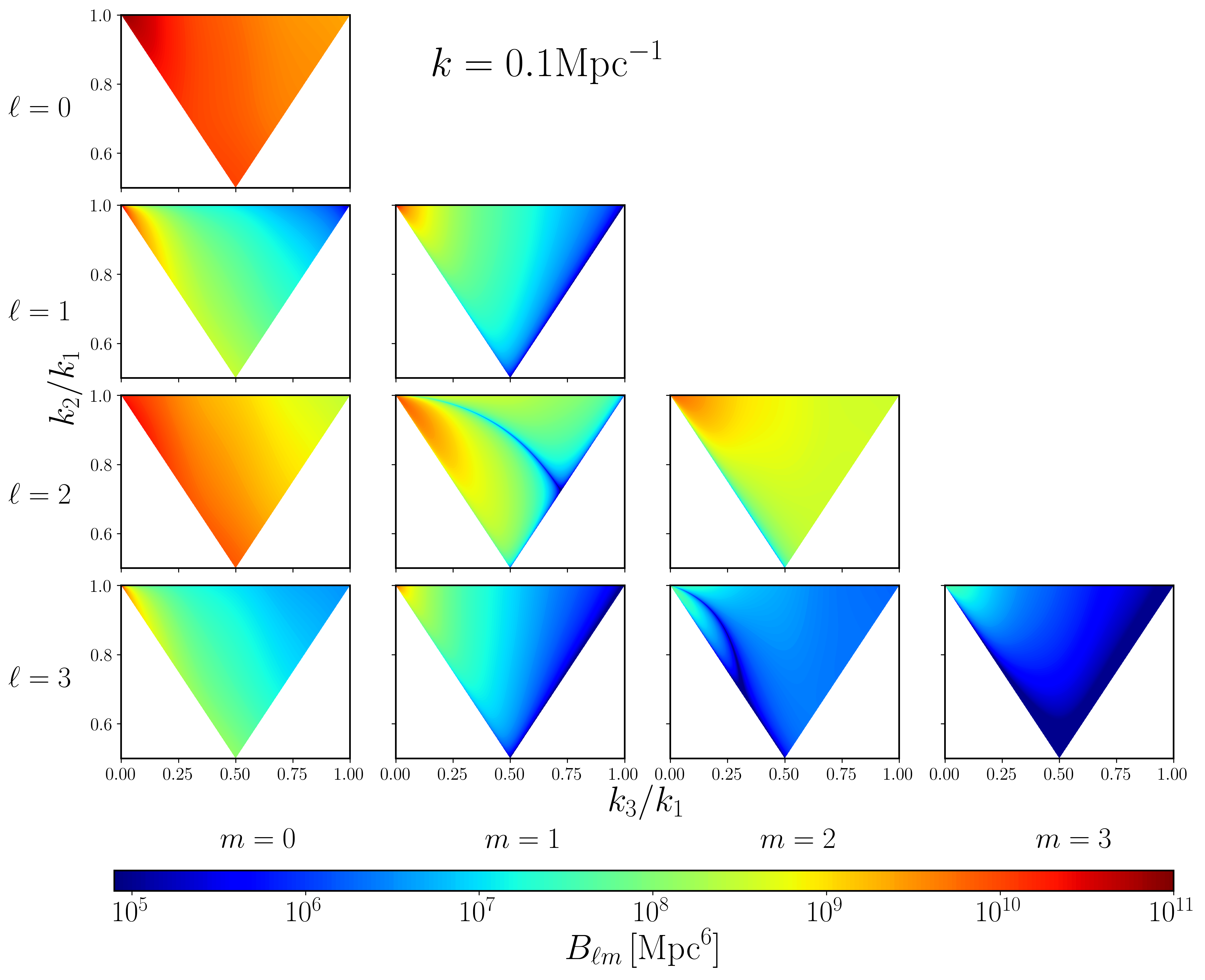}
	\caption{A selection of multipoles of the galaxy bispectrum, \(B_{\ell m}\), with \(\ell = 0 \ldots 3\) and \(m = 0 \ldots \ell\) as indicated in the figure. Bias model used is that for \(H\alpha\)/Euclid-like survey. \(k_1\) is kept fixed, the value of which is given alongside the plot, and the \(x\) and \(y\) axes vary respectively \(k_3\) and \(k_2\) with respect to the fixed \(k_1\). The upper left corner of the wedge shape is the squeezed limit, the upper right corner is the equilateral configuration, and the lower corner is the co-linear configuration. Note the difference in range of the colour bars.}
		\label{fig:euclidtriangle}
\end{figure}

\clearpage
\begin{figure}[H]
	\centering
	\includegraphics[width=0.8\textwidth]{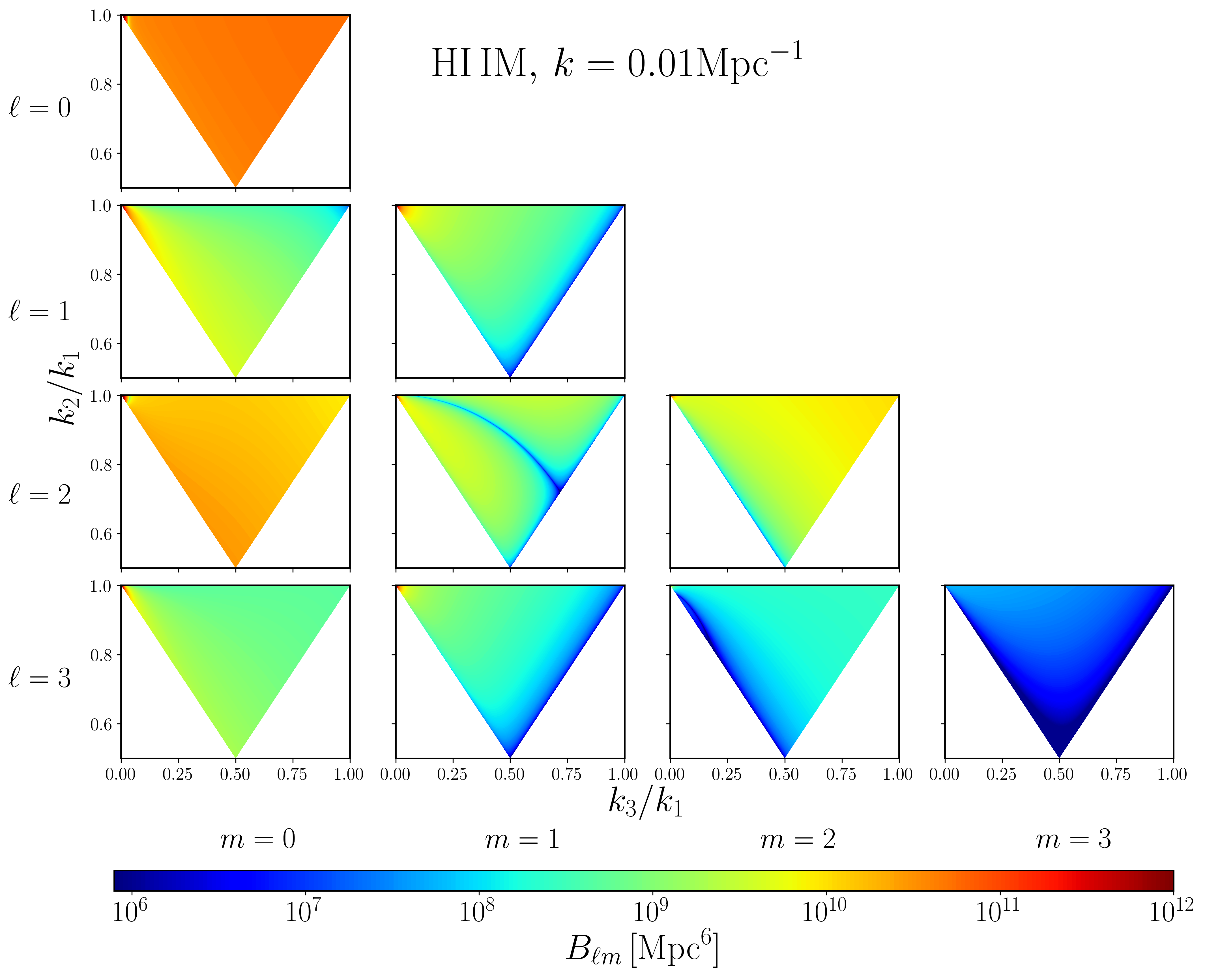}
	\includegraphics[width=0.8\textwidth]{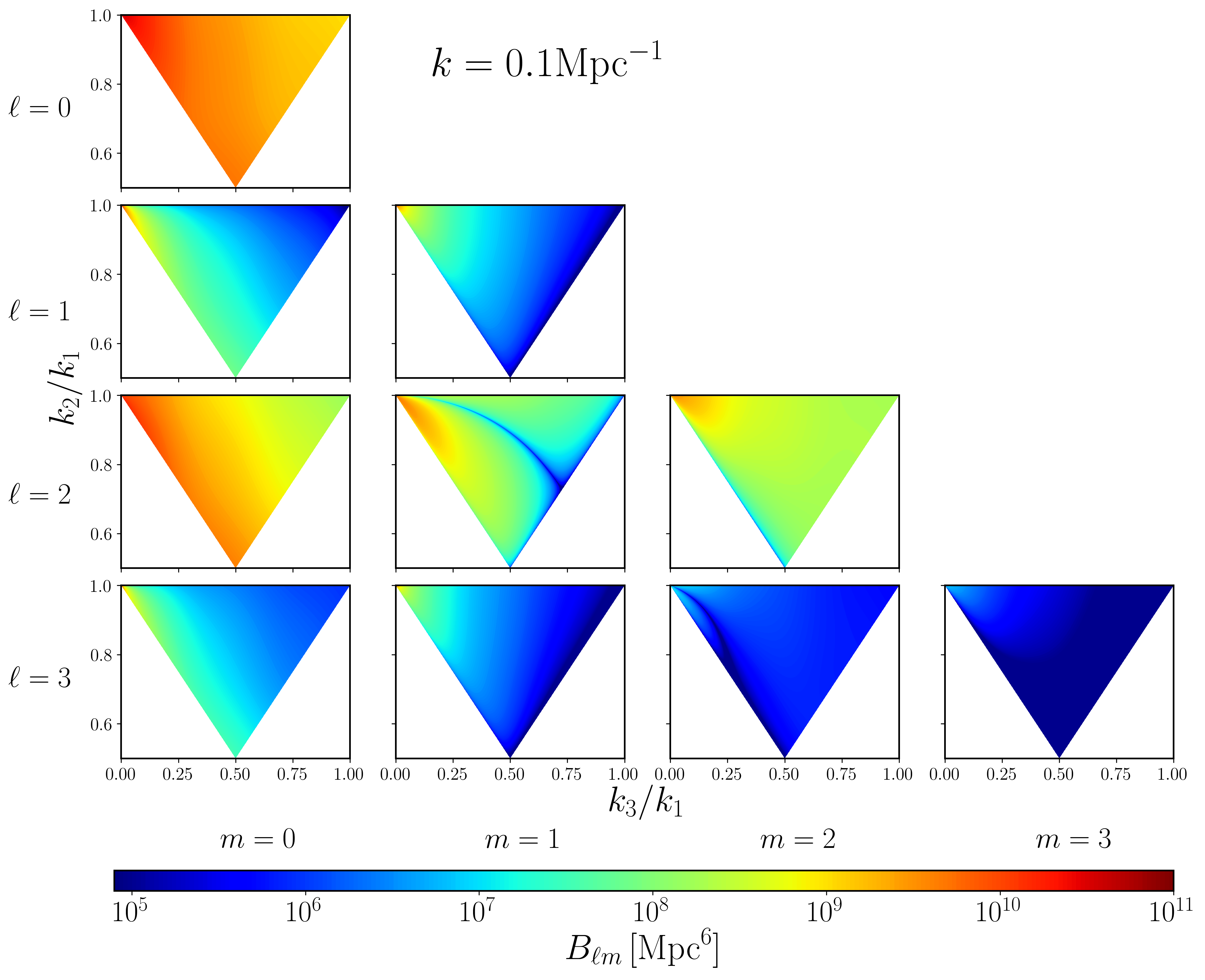} 
	\caption{Selected multipoles of the galaxy bispectrum, similar to figure~\ref{fig:euclidtriangle}, but with the bias model appropriate for intensity mapping. The value of fixed \(k_1\) is indicated on the figures. The upper left corner of a wedge shape is the squeezed limit, the upper right corner is the equilateral configuration, and the lower corner is the co-linear configuration. \label{fig:imtriangle}}
\end{figure}

\clearpage
\begin{figure}[H]
	\centering
	\includegraphics[width=0.85\linewidth]{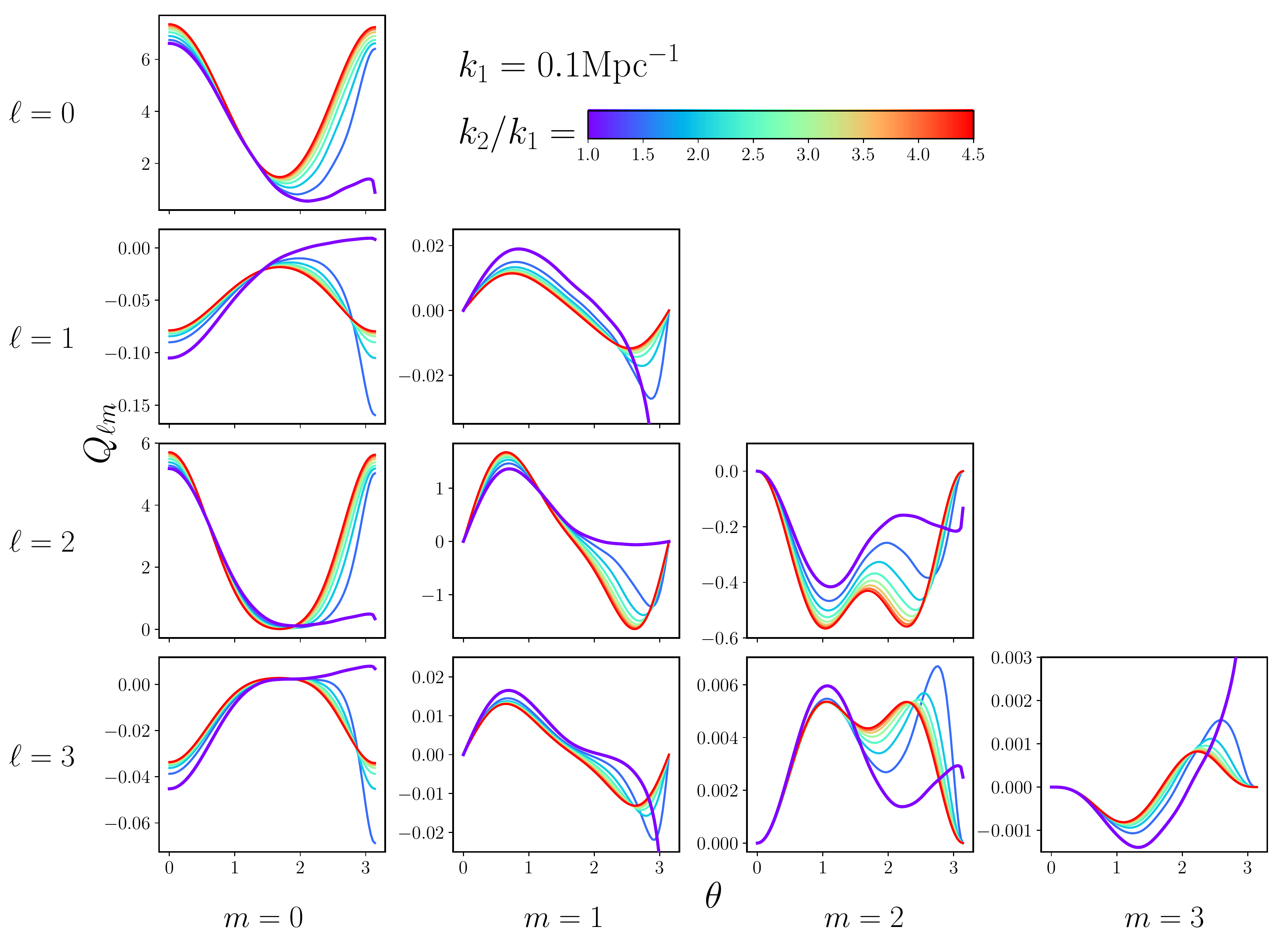}
	\includegraphics[width=0.85\linewidth]{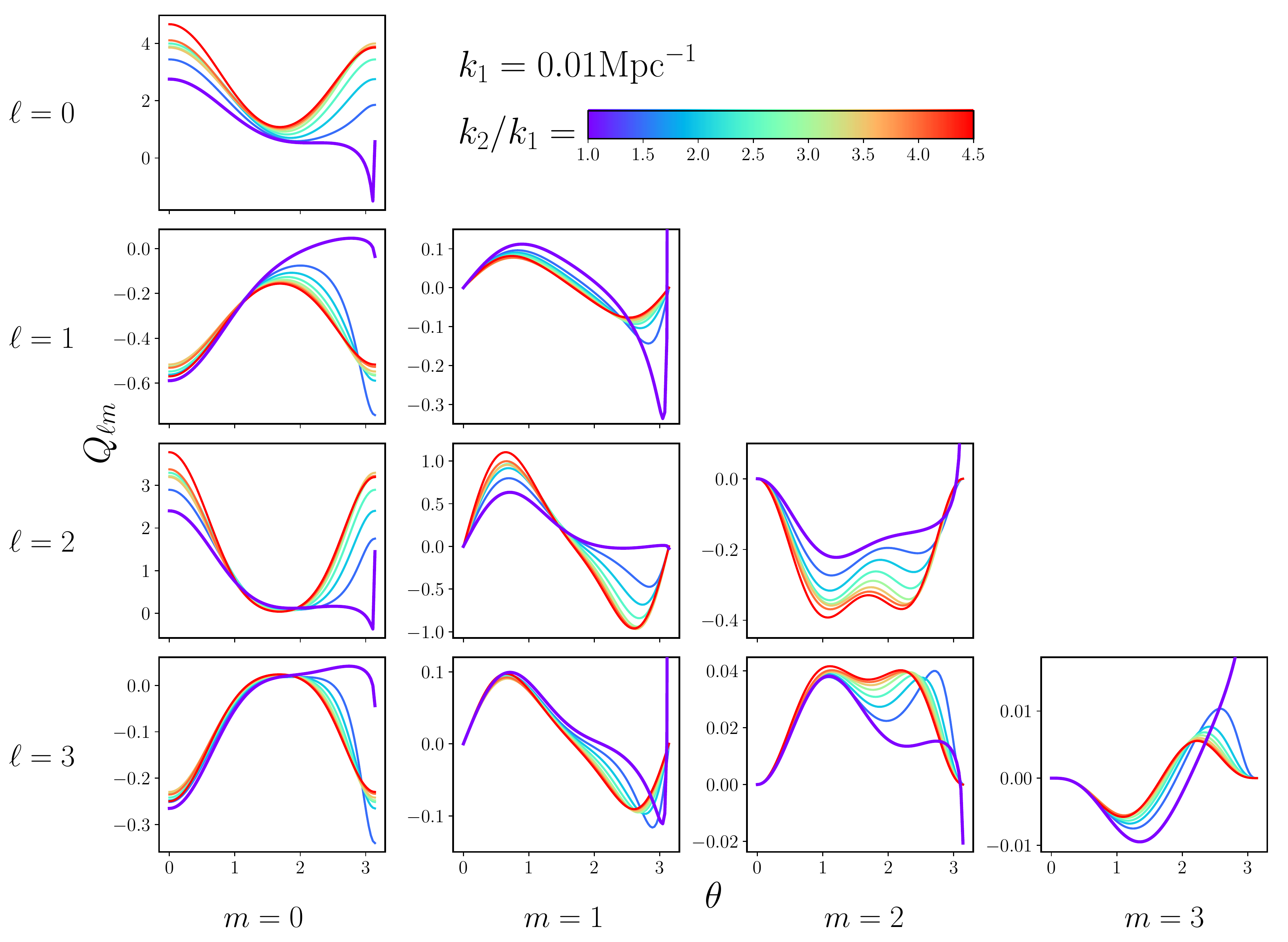}
	\caption{Results for the reduced bispectrum \(Q_{\ell m}\), where \(\ell = 0 \ldots 3\) and negative \(m\) not shown. The multipoles \(\ell,m\) are indicated on the figure, as well as the value of \(k_1\) which is kept fixed. The colourbar and different colours denote the ratio of \(k_2/k_1\), where the slightly thicker purple line is the isosceles triangle, which diverges as \(\theta\to\pi\) since there \(k_3\to 0\). \label{fig:Q_rb}}
\end{figure}

\clearpage
\begin{figure}[H]
\centering
\includegraphics[width=\linewidth]{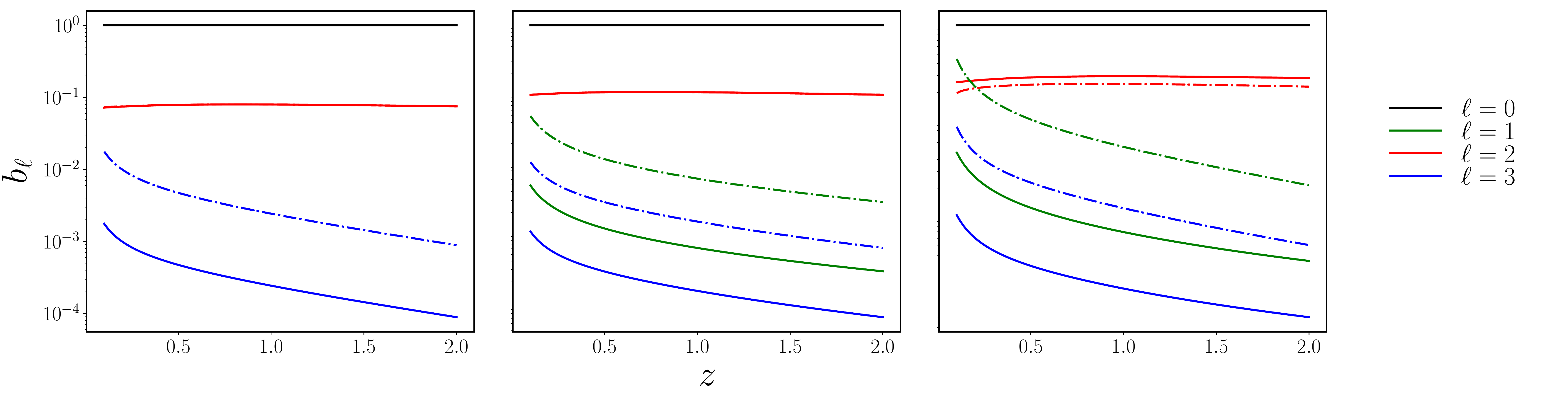}
\caption{Total power contained in the relativistic bispectrum normalised by the Newtonian monopole as a function of redshift \(z\) ranging from 0.1 to 2. The three panels are the equilateral configuration (left, with \(\ell = 1\) vanishing), squeezed (middle) and co-linear flattened configuration (right). Solid lines for \(k = 0.1 \, \mathrm{Mpc^{-1}}\), and dash-dotted for \(k = 0.01 \, \mathrm{Mpc^{-1}}\).\label{fig:fnofz}}
\end{figure}

Figures~\ref{fig:totpowersq},\ref{fig:totpowerfl} and~\ref{fig:totpowereq} show the amplitude of the total power as defined in~\eqref{eq:totpower}. For each $\ell$ this contains all orientations per multipole divided by the amplitude of the Newtonian monopole. Values of \(\ell\) are labelled on the figure, with the dotted lines denoting the Newtonian contribution (for even \(\ell\) only). For small scales (larger wavenumber \(k\)), the Newtonian contributions are generally larger than the relativistic \(b_\ell\) (i.e. odd \(\ell\)), however at larger scales,  above equality, the relative power contained in relativistic contributions increases. This shows up in the even multipoles as a divergence between the dotted (purely Newtonian) lines and solid (GR-corrected) lines. In the odd multipoles, we see an increase in amplitude, which at the largest scales become larger than the purely Newtonian signal. This is dependent on bias model and triangular configuration.

The colour-intensity maps in figures~\ref{fig:euclidtriangle} and~\ref{fig:imtriangle} show the amplitude of the relativistic bispectrum over the \(k_3/k_1\), \(k_2/k_1\) plane. The amplitude of the bispectrum signal peaks in the squeezed limit where \(k_1=k_2\), \(k_3\to 0\) which is in the top left corner in these plots. For the odd multipoles \(\ell = 1\) and \(\ell = 3\),  the amplitude of the dipole is higher than the \(\ell=3\) case in most configurations. The amplitude of the relativistic bispectrum is also higher for larger scales (smaller \(k\)). For \(\ell = 1\), the equilateral configuration, which lies in the upper right corner of the plots, is vanishing as we established analytically. 
We can also observe from these plots that there is a rough trend that more power is contained in the lower $m$ multipoles.

The reduced bispectrum is plotted in figure~\ref{fig:Q_rb}, showing large relativistic contributions to the bispectrum odd-multipoles especially at large scales. This also shows the significant dependence on the triangle shape, depending on the orientation of the harmonic.  

Finally figure~\ref{fig:fnofz} shows the total power divided by the Newtonian monopole, as a function of redshift. The model for bias used here is not physically realistic, but this illustrates the generic behaviour with redshift we can expect. It is interesting to observe how, when going towards lower redshift, the power in the relativistic corrections to the bispectrum grows compared to the Newtonian signal. This is especially noticeable in squeezed and flattened shapes where the dipole approaches or surpasses the \(\ell = 2\) line. Of course, at low redshift the plane-parallel assumption that we have used becomes a worse approximation.

\section{Conclusion}\label{sec:concl}

We have considered in detail for the first time the multipole decomposition of the observed relativistic galaxy bispectrum. In section~\ref{sec:extrmulti} we have shown how the multipoles may be derived analytically, with an analytic formula given in equation~\eqref{eq:sumformula}, and have illustrated how they behave in the squeezed, equilateral and co-linear limits (which includes the flattened case) in section~\ref{sec:anal}. We have shown how the amplitude of the relativistic signals behaves for two types of upcoming surveys~-- a Euclid-like galaxy survey, and an SKA intensity mapping survey. Our key findings are:
\begin{description}
\item[odd multipoles] Relativistic effects generate a hierarchy of odd multipoles which are absent in the Newtonian picture, plus an additional contribution to all multipoles up to $\ell=7$. In particular we find that the octopole is similar in amplitude to the dipole; it is only about a factor of 5 or so smaller than the dipole. These are both larger than the Newtonian hexadecapole on large scales. Higher multipoles are suppressed. This effect can be seen clearly in figures~\ref{fig:totpowersq},~\ref{fig:totpowerfl},~\ref{fig:totpowereq}.
\item[powers of $k$] The leading power of the relativistic correction in each $\ell$ harmonic is $(\cH/k)^1$ for odd multipoles and $(\cH/k)^2$, for even multipoles. Furthermore, all odd multipoles contain the leading $(\cH/k)$ correction, while lower values of \(\ell\) contain the higher powers of \(\cH/k\), going up to \((\cH/k)^7\) for \(\ell=1\) (though these are probably unobservable). An overview of occurring powers of \(k\) is given in figure~\ref{fig:blm_overview}.
\item[special limits] the co-linear case ($\theta=0$ or $\pi$) only generates non-zero $m=0$ multipoles and vanishes for all other values of \(m\). The equilateral case is always zero for $m$ odd, and is always zero for the special case of the dipole. For the squeezed limit we have leading $(\cH/k)$ relativistic corrections for $\ell$ and $m\leq3$ odd.
\item[multipoles with shape] We computed the amplitude of each $\ell,m$ over the range of triangle shapes in figures~\ref{fig:euclidtriangle},~\ref{fig:imtriangle}. For each $\ell$ most of the power is contained in the lower $m$ multipoles. 
\item[multipoles with scale] We analysed the total power in each multipole as a function of scale for 3 triangle shapes at $z=1$. Roughly speaking the even-$\ell$ are dominated by the Newtonian part and have little scale dependence relative to the Newtonian monopole, though this changes approaching the Hubble scale. For odd-$\ell$ the leading relativistic part dominates and the dipole reaches the size of the Newtonian quadrupole around equality scales.
%
\item[redshift dependence] Relative to the Newtonian monopole, all the relativistic multipoles decay with redshift, while the quadrupole is roughly constant. For large squeezed triangles the dipole is comparable in size to the quadrupole for small redshift as shown in figure~\ref{fig:fnofz}.
\end{description}
Of course, the analysis here is limited by the fact we have neglected wide angle effects which will alter the multipoles.  Integrated effects will also contribute, but their effect will be suppressed when we analyse the multipoles. We leave these contributions for future work. Also currently under investigation is detectability of the galaxy bispectrum, with the leading order contribution examined in~\cite{Maartens:2019yhx}.

\appendix

\section{Beta coefficients}
\label{sec:betacoeffappendix}
Here we list the time and bias dependent coefficients appearing in the relativistic second-order kernel. 
\begin{align} 
\frac{\beta_{1}}{\mathcal{H}^{4}} &= \frac{9}{4}\Omega_{m}^{2}\Bigg[6-2f\bigg(2b_{e}-4\Q-\frac{4(1-\Q)}{\chi\cH}-\frac{2\cH'}{\cH^{2}}\bigg)-\frac{2f'}{\cH}+b_{e}^{2}+5b_{e}-8b_{e}\Q + 4\Q + 16\Q^{2} \nonumber\\ 
& - 16\frac{\p \Q}{\p \ln{\bar{L}}} - 8\frac{\Q'}{\cH} + \frac{b_{e}'}{\cH}+\frac{2}{\chi^{2}\cH^{2}}\bigg(1-\Q+2\Q^{2}-2\frac{\p \Q}{\p \ln{\bar{L}}}\bigg)
 \nonumber\\ 
& - \frac{2}{\chi\cH}\bigg(3+2b_{e}-2b_{e}\Q-3\Q +8\Q^{2}-\frac{3\cH'}{\cH^{2}}(1-\Q) -8\frac{\p \Q}{\p \ln{\bar{L}}} - 2\frac{\Q'}{\cH}\bigg) \nonumber \\
& + \frac{\cH'}{\cH^{2}}\bigg(-7-2b_{e}+8\Q+\frac{3\cH'}{\cH^{2}}\bigg) - \frac{\cH''}{\cH^{3}}\Bigg] \nonumber \\
&  +\frac{3}{2}\Omega_{m}f\Bigg[5-2f(4-b_{e})+\frac{2f'}{\cH}+2b_{e}\bigg(5+\frac{2(1-\Q)}{\chi\cH}\bigg)-\frac{2b_{e}'}{\cH} -2b_{e}^{2} + 8b_{e}\Q - 28\Q \nonumber \\
& - \frac{14(1-\Q)}{\chi\cH}-\frac{3\cH'}{\cH^{2}} +4\bigg(2-\frac{1}{\chi\cH}\bigg)\frac{\Q'}{\cH}\Bigg] \nonumber \\
&  +\frac{3}{2}\Omega_{m}f^{2}\bigg[-2+2f-b_{e}+4\Q+\frac{2(1-\Q)}{\chi\cH}+\frac{3\cH'}{\cH^{2}}\bigg] \nonumber \\
&  +f^{2}\bigg[12-7b_{e}+b_{e}^{2}+\frac{b_{e}'}{\cH}+(b_{e}-3)\frac{\cH'}{\cH^{2}}\bigg] - \frac{3}{2}\Omega_{m}\frac{f'}{\cH} \\
\nonumber\\
\frac{\beta_{2}}{\cH^{4}} &= \frac{9}{2}\Omega_{m}^{2}\bigg[-1+b_{e}-2\Q-\frac{2(1-\Q)}{\chi\cH}-\frac{\cH'}{\cH^{2}}\bigg] + 3\Omega_{m}f\bigg[-1+2f{-b_{e}}+4\Q+\frac{2(1-\Q)}{\chi\cH}+\frac{3\cH'}{\cH^{2}}\bigg] \nonumber \\
&  +3\Omega_{m}f^{2}\bigg[-1+b_{e}-2\Q-\frac{2(1-\Q)}{\chi\cH}-\frac{\cH'}{\cH^{2}}\bigg] + 3\Omega_{m}\frac{f'}{\cH}\; \\ 
\frac{\beta_{3}}{\mathcal{H}^{3}} &= \frac{9}{4}\Omega_{m}^{2}(f-2+2\Q) \nonumber \\
&  + \frac{3}{2}\Omega_{m}f\Bigg[-2 -f\bigg(-3+f+2b_{e}-3\Q-\frac{4(1-\Q)}{\chi\cH}-\frac{2\cH'}{\cH^{2}}\bigg)-\frac{f'}{\cH}\nonumber \\
& +{3}b_{e}+b_{e}^{2}-6b_{e}\Q{+4}\Q +8\Q^{2}-8\frac{\p \Q}{\p \ln{\bar{L}}} -6\frac{\Q'}{\cH} +\frac{b_{e}'}{\cH} \nonumber \\
& +\frac{2}{\chi^{2}\cH^{2}}\bigg(1-\Q+2\Q^{2}-2\frac{\p \Q}{\p \ln{\bar{L}}}\bigg) + \frac{2}{\chi \cH}\bigg({-1} -2b_{e}+2b_{e}\Q{+}\Q-6\Q^{2}  \nonumber \\
& +\frac{3\cH'}{\cH^{2}}(1-\Q) +6\frac{\p \Q}{\p \ln{\bar{L}}} + 2\frac{\Q'}{\cH}\bigg) -\frac{\cH'}{\cH^{2}}\bigg({3}+2b_{e}-6\Q-\frac{3\cH'}{\cH^{2}}\bigg) - \frac{\cH''}{\cH^{3}}\Bigg] \nonumber \\
&  {+} f^{2}\Bigg[-3+2b_{e}\bigg(2+\frac{(1-\Q)}{\chi\cH}\bigg)-b_{e}^{2}+2b_{e}\Q -6\Q-\frac{b_{e}'}{\cH}-\frac{6(1-\Q)}{\chi\cH}\nonumber \\
& +2\bigg(1-\frac{1}{\chi\cH}\bigg)\frac{\Q'}{\cH}\Bigg]  \\ 
\frac{\beta_{4}}{\cH^{3}} &= \frac{9}{2}\Omega_{m}f\bigg[-b_{e}+2\Q+\frac{2(1-\Q)}{\chi \cH}+\frac{\cH'}{\cH^{2}}\bigg]\\ 
{{\frac{\beta_5}{\cH^{3}}}} &= 3\Omega_{m}f\bigg[b_{e}-2\Q-\frac{2(1-\Q)}{\chi\cH}-\frac{\cH'}{\cH^{2}}\bigg] \\ 
\frac{\beta_6}{\cH^{2}} &= \frac{3}{2}\Omega_{m}\Bigg[2-2f+b_{e}-4\Q-\frac{2(1-\Q)}{\chi\cH}-\frac{\cH'}{\cH^{2}}\Bigg] \\
\frac{\beta_7}{\cH^{2}} &= f(3-b_{e}) \\ 
\frac{\beta_8}{\cH^{2}} &= {3\Omega_{m}f(2-f-2\Q)} + f^{2}\Bigg[4+b_{e}-b_{e}^{2}+4b_{e}\Q-{6}\Q-4\Q^{2}+4\frac{\p \Q}{\p \ln{\bar{L}}} + 4\frac{\Q'}{\cH} - \frac{b_{e}'}{\cH}  \nonumber \\\nonumber\\
& - \frac{2}{\chi^{2}\cH^{2}}\bigg(1-\Q+2\Q^{2}-2\frac{\p \Q}{\p \ln{\bar{L}}}\bigg) - \frac{2}{\chi\cH}\bigg(3-2b_{e}+2b_{e}\Q-\Q-4\Q^{2}+\frac{3\cH'}{\cH^{2}}(1-\Q) \nonumber\\
& + 4\frac{\p \Q}{\p \ln{\bar{L}}} + 2\frac{\Q'}{\cH}\bigg) - \frac{\cH'}{\cH^{2}}\bigg(3-2b_{e}+{4}\Q+\frac{3\cH'}{\cH^{2}}\bigg) + \frac{\cH''}{\cH^{3}}\Bigg] \\ 
\frac{\beta_{9}}{\cH^{2}} &= -\frac{9}{2}\Omega_{m}f \\ 
{\frac{\beta_{10}}{\cH^{2}}} &= 3\Omega_{m}f \\ 
\frac{\beta_{11}}{\cH^{2}} &= 3\Omega_{m}\bigg(\frac{1}{2}+f\bigg) + f - f^{2}\Bigg[-1+b_{e}-2\Q- \frac{2(1+\Q)}{\chi\cH}-\frac{\cH'}{\cH^{2}}\Bigg] \\ 
\frac{\beta_{12}}{\cH^{2}} &= \frac{3}{2}\Omega_{m}\Bigg[-2+b_{1}\bigg(2+b_{e}-4\Q-\frac{2(1-\Q)}{\chi\cH} -\frac{\cH'}{\cH^{2}}\bigg) + \frac{b_{1}'}{\cH} + 2\bigg(2-\frac{1}{\chi\cH}\bigg)\frac{\p b_{1}}{\p \ln{\bar{L}}}\Bigg] \nonumber \\
&- f\Bigg[2+b_{1}(f-3+b_{e}) + \frac{b_{1}'}{\cH}\Bigg]  \\ 
\frac{\beta_{13}}{\cH^{2}} &= \frac{9}{4}\Omega_{m}^{2} {+ \frac{3}{2}\Omega_{m}f\Bigg[1-2f+2b_{e}-{6}\Q-\frac{4(1-\Q)}{\chi\cH}-\frac{3\cH'}{\cH^{2}}\Bigg]} + f^{2}(3-b_{e}) \\ 
\frac{\beta_{14}}{\cH} &= -\frac{3}{2}\Omega_{m}b_{1} \\
\frac{\beta_{15}}{\cH} &= 2f^{2}  \\ 
\frac{\beta_{16}}{\cH} &= f\Bigg[b_{1}\bigg(f+b_{e}-2\Q-\frac{2(1-\Q)}{\chi\cH}-\frac{\cH'}{\cH^{2}}\bigg) + \frac{b_{1}'}{\cH} + 2\bigg(1-\frac{1}{\chi\cH}\bigg)\frac{\p b_{1}}{\p \ln \bar{L}}\Bigg] \\
\frac{\beta_{17}}{\cH} &= -\frac{3}{2}\Omega_{m}f \\
\frac{\beta_{18}}{\cH} &= \frac{3}{2}\Omega_{m}f -f^{2}\Bigg[3-2b_{e}+{4}\Q+\frac{4(1-\Q)}{\chi\cH}+\frac{3\cH'}{\cH^{2}}\Bigg] \\
\frac{\beta_{19}}{\cH} &= f\Bigg[b_{e}-2Q-\frac{2(1-\Q)}{\chi\cH}-\frac{\cH'}{\cH^{2}}\Bigg] 
\end{align}

\section{Derivation of the sum formula}\label{sec:derivationsumappendix}
Here we present the derivation of the analytic result \ref{eq:sumformula}, that is, exact integration of:
\begin{equation} X_{\ell m}^{ab} = \int_0^{2\pi} \diff\varphi\, \int_{-1}^1 \diff\mu_1\, (\i \mu_1)^a (\i \mu_2)^b \,Y_{\ell m}^*(\mu_1,\varphi).
\end{equation}
We will calculate this for $m\geq0$, as for negative $m$ we can use the result
\begin{equation}
X_{\ell, -m}^{ab} = (-1)^{a+b+m} X_{\ell m}^{ab*}\,,
\end{equation}
which follows on using the complex conjugate of the standard orthonormal spherical harmonics,
\begin{align}
Y_{\ell m}^* &= (-1)^m Y_{\ell, -m} \nonumber\\
&= \sqrt{\frac{2 \ell + 1}{4 \pi}} \sqrt{\frac{(\ell - m )!}{(\ell + m )!}} P_\ell^m(\mu) e^{-\i m \varphi}.
\end{align}

To perform this integral analytically, first use the binomial expansion to expand the \(\mu_i\) dependence in the integrand, 
\begin{equation}
( \i \mu_1)^a \left(\i \sqrt{1-\mu_1^2} \sin\theta \sin\varphi + \i \mu_1 \cos\theta \right)^b,
\end{equation}
as
\begin{align}
(\i \mu_2)^b &= \i^b \sum_{g=0}^b \binom{b}{g} (\mu_1 \cos\theta)^{b-g} \left(\sqrt{1-\mu_1^2}\sin\theta \sin\varphi\right)^g \nonumber\\
& =\i^b \sum_{g=0}^b \binom{b}{g} \mu_1^{b-g} \left(\sqrt{1-\mu_1^2}\right)^g \cos\theta^{b-g} \sin\theta^g \sin\varphi^g. 
\end{align}
Now the separability of the angular parts of the integrand has been made explicit. Inserting this expansion backinto the integral we get, 
\begin{align}
&\sum_{g=0}^b \i^{a+b} \cos^{b-g}\theta \sin^g\theta \binom{b}{g} \int_0^{2\pi}\diff\varphi\, \int_{-1}^1 \diff\mu_1\, \mu_1^{a+b-g} \left(1-\mu_1^2\right)^{g/2} \sin^g\varphi Y_{\ell m}^*(\mu_1,\varphi),
\end{align}
where the factors that are independent of integration angles \(\mu_1,\,\varphi\) have been taken out of the integral (note that \(\theta = \theta_{12}\) as per our convention used throughout this paper). 

In what follows will drop the subscript on \(\mu_1 = \mu\) for convenience. Using the standard definition of the spherial harmonics, the integral then becomes,
\begin{equation}
	\sum_{g=0}^b \i^{a+b} \cos^{b-g}\theta \sin^g\theta \binom{b}{g} \sqrt{\frac{2\ell+1}{4 \pi}} \sqrt{\frac{(\ell-m)!}{(\ell+m)!}}\int_0^{2\pi}\diff\varphi\, \int_{-1}^1 \diff\mu\, \mu^{a+b-g} (1-\mu^2)^{g/2} \sin^g\varphi P_\ell^m(\mu) e^{-\i m \varphi},
\end{equation}
and hence can easily be split into two parts. The associated Legendre polynomials \(P_\ell^m\) can be expressed as 
\begin{equation}
P_\ell^m(\mu) = (-1)^m (1-\mu^2)^{m/2} \frac{\diff^m}{\diff\mu^m} P_\ell(\mu), 
\end{equation}
i.e. as full derivatives of the Legendre polynomials. These in turn can be expressed as a sum 
\begin{equation}
P_\ell(\mu) =2^\ell \sum_{h=0}^\ell \mu^\ell \binom{\ell}{h} \binom{\frac{1}{2}(\ell+h-1)}{\ell}.
\end{equation}
Using the Legendre polynomials in this form and substituting in,
\begin{align}
	&\int_{-1}^1 \diff\mu\, \mu^{a+b-g} \left(1-\mu^2\right)^{g/2} P_\ell^m(\mu) \nonumber \\
	&= (-1)^m 2^\ell \sum_{h=0}^\ell \binom{\ell}{h} \binom{\frac{1}{2}(\ell+h-1)}{\ell} \int_{-1}^1 \diff\mu\, \mu^{a+b-g} \left(1-\mu^2\right)^{g/2}  (1-\mu^2)^{m/2} \frac{\diff^m}{\diff\mu^m} \mu^\ell \nonumber \\
	&= (-1)^m 2^\ell \sum_{h=0}^\ell \binom{\ell}{h} \binom{\frac{1}{2}(\ell+h-1)}{\ell} \frac{h!}{(h-m)!} \frac{1}{2} \left(1+ (-1)^{a+b-g+h-m}\right) \Gamma\left[\frac{1}{2}(a+b-g+h-m-1)\right] \Gamma\left[\frac{1}{2} (g+m+2)\right] \nonumber \\
	&\times  \left\{ \Gamma\left[\frac{1}{2} (a+b+h+3)\right] \right\}^{-1},
\end{align}
where \(m \geq h\), so that above result may be written as
\begin{align}
	&(-1)^m 2^\ell \sum_{h=m}^\ell \binom{\ell}{h} \binom{\frac{1}{2}(\ell+h-1)}{\ell} \frac{h!}{(h-m)!} \frac{1}{2} \left(1+ (-1)^{a+b-g+h-m}\right) \Gamma\left[\frac{1}{2}(a+b-g+h-m-1)\right] \Gamma\left[\frac{1}{2} (g+m+2)\right] \nonumber \\
	&\times  \left\{ \Gamma\left[\frac{1}{2} (a+b+h+3)\right] \right\}^{-1}.
\end{align}
Evaluating now the integral over \(\varphi\), which is, 
\begin{align}
	&\int_0^{2\pi} \diff\varphi\, \sin^g \varphi e^{-\i m \varphi} \nonumber \\
	& = \frac{1}{(2 \i)^g} \sum_{n=0}^g \binom{g}{n} (-1)^n \int_0^{2\pi}\diff\varphi\,  e^{\i (g-m-2n )\varphi} \nonumber \\
	&= \frac{1}{(2 \i)^g}  \sum_{n=0}^g \binom{g}{n} (-1)^n (2 \pi) \delta_{g-m-2n, 0}.
\end{align}

The Kronecker \(\delta\) picks out one of the terms in the sum, \(g-m-2n=0 \to n = \frac{g-m}{2}\), so 
\begin{equation}
	\int_0^{2\pi} \diff\varphi\, \sin^g \varphi e^{-\i m \varphi} = \frac{1}{(2 \i )^g} \binom{g}{\frac{1}{2}(g-m)} 2 \pi (-1)^{\frac{g-m}{2}},
\end{equation}
if \(g-m\) is even, in which case \((-1)^{\frac{g-m}{2}} = 1\), so
\begin{equation}
	\int_0^{2\pi} \diff\varphi\, \sin^g \varphi e^{- \i m \varphi} = 2^{-g} \i^g (-1)^g \binom{g}{\frac{1}{2}(g-m)} 2 \pi, 
\end{equation}
for \(g+m \) even, zero otherwise. 

Putting the results from both integrals together, 
\begin{align}
	&\int_0^{2\pi} \diff\varphi\, \int_{-1}^1 \diff\mu_1\, (\i \mu_1)^a (\i \mu_2)^b \,Y_{\ell m}^*(\mu_1,\varphi) = \sum_{g=0}^b \i^{a+b} \cos^{b-g}\theta \sin^g\theta \binom{b}{g} \sqrt{\frac{2\ell+1}{4 \pi}} \sqrt{\frac{(\ell-m)!}{(\ell+m)!}} \times \left\{(-1)^m 2^\ell \sum_{h=m}^\ell \binom{\ell}{h} \right. \nonumber \\
	&\left. \binom{\frac{1}{2}(\ell+h-1)}{\ell} \frac{h!}{(h-m)!}  \frac{1}{2} \left(1+ (-1)^{a+b-g+h-m}\right) \Gamma\left[\frac{1}{2}(a+b-g+h-m-1)\right] \Gamma\left[\frac{1}{2} (g+m+2)\right]  \times \right. \nonumber \\
	&\left.  \left[\Gamma\left[\frac{1}{2} (a+b+h+3)\right] \right\}^{-1} \right]\times \left\{  2^{-g} \i^g (-1)^g \binom{g}{\frac{1}{2}(g-m)} 2 \pi\right\}.
\end{align}
Simplifying the above result,
\begin{align}
	&\sum_{p=m}^{\frac{1}{2}(b+m)} \i^{a+b+m} \cos^{b-2p+m}\theta \sin^{2p-m}\theta \frac{b!}{(2p-m)! (b-2p+m)!} \sqrt{\frac{\pi (2\ell+1) (\ell-m)!}{(\ell+m)!}} \sum_{q=m}^\ell 2^\ell \frac{\ell!}{q! (\ell-q)!} \nonumber \\
	& \frac{(\frac{1}{2}(\ell+q-1))!}{\ell!(\frac{1}{2}(\ell+q-1)-l)!} \frac{q!}{(q-m)!} 2^{-1} (1+(-1)^{a+b+q}) \Gamma\left[\frac{1}{2} (a+b-2p+q+1)\right] \Gamma\left[\frac{1}{2}(2p+2)\right]\times \nonumber \\
	& \frac{(2p-m)!}{(\frac{1}{2}(2p-2m))! (2p-m - \frac{1}{2}(2p-2m))!} \times \left\{ \Gamma\left[\frac{1}{2}(a+b+q+3)\right] \right\}^{-1},
\end{align}
collecting terms, and after cancellations some cancellations obtain, 
\begin{align}
	&\sum_{p=m}^{\frac{1}{2}(b+m)} \sum_{q=m}^\ell 2^{\ell+m-1} \i^{a+b+m} \cos^{b-2p+m}\theta \sin^{2p-m}\theta  \sqrt{\frac{\pi (2\ell+1) (\ell-m)!}{(\ell+m)!}} b! \left(\frac{1}{2}(\ell+q-1)\right)! \left(\frac{1}{2}(a+b-2p+q-1)\right)! \times \nonumber \\
	& \left(1+(-1)^{a+b+g}\right) \times \left[ 4^p (p-m)! (b+m-2p)! (\ell-q)! \left(\frac{1}{2}(-\ell+q-1)\right)! (q-m)! \left(\frac{1}{2}(a+b+q+1)\right)! \right]^{-1},
\end{align}
where we have used \(\Gamma(n)=(n-1)!\) to rewrite the gamma functions in terms of factorials, that this is non-zero only if \(\frac{1}{2}(g-m)\) is even, and that \((-1)^{w}=1\) if \(w\) is even. The final analytic expression for \(m>0\) hence is,  
\begin{align}
\int_0^{2\pi}{\rm d}\varphi\,& \int_{-1}^{1}{\rm d}\mu_1\,(\i\mu_1)^a\left(\i\sqrt{1-\mu_1^2} \sin\theta \sin\varphi + \i\mu_1\cos\theta\right)^b \, Y^*_{\ell m} (\mu_1,\varphi)  = 2^{\ell+m-1}\i^{a+b+m}\sqrt{\frac{\pi(2\ell+1)(\ell-m)!}{(\ell+m)!}} \nonumber \\
&\times
\sum_{p=m}^{\frac{1}{2}(b+m)}\sum_{q=m}^{\ell}
\frac{\left[1+(-1)^{a+b+q}\right]\,b!\,\cos^{b+m-2p}\theta\sin^{2p-m}\theta}{4^p(b+m-2p)!(\ell-q)!(p-m)!(q-m)!}
\frac{\Gamma\left[\frac{1}{2}(q+\ell+1)\right]}{\Gamma\left[\frac{1}{2}(q-\ell+1)\right]}
\frac{\Gamma\left[\frac{1}{2}(a+b +q-2p+1)\right]}{\Gamma\left[\frac{1}{2}(a+b+q+3)\right]}.
\end{align}
Note that in the above we have kept the expression in terms of gamma functions, but this can easily be reverted back to the factorial notation.

\section{\(\ko_{ab}^{(n)}\) kernel coefficients}\label{sec:kabappendix}

Here we present the higher order \(\cH/k\) kernels for the first of the cyclic permutations. It is worth noting that these cannot be exacly manipulated to obtain the coefficients for the other two cyclic permutations, since making the replacements \(\mu_1 \to \mu_2, \, \mu_3\) introduces additional powers of \(\mu_i\), giving rise to slightly different coefficients \(\ko_{ab}\). It is however easy enough to extract the coefficients for these permutations following the same method. Below we focus on only the first of the cyclic permutations, that is, the 123 permutation, as outlined before. Schematic representations of the higher order Newtonian and GR kernels are given, along with their corresponding coefficients. Like before, for brevity we use shorthand notations; \(F = F_2(\ka,\kb)\), \(G = G_2(\ka,\kb)\), and \(S = S_2(\ka,\kb)\). Superscript \(n\) on \(\ko_{ab}^{(n)}\) denotes the power \((\cH/k)^n\).

\begin{equation}
\ko_{ab}^{(2)} =  \left( \begin {array}{cccccc} \bullet &\circ &\bullet &\circ 
&\bullet &\circ \\  \circ &\bullet &\circ &
\bullet &\circ &\bullet \\  \bullet &\circ 
&\bullet &\circ &\bullet &\circ \\  \circ &
\bullet &\circ &\bullet &\circ &\circ \\  
\bullet &\circ &\bullet &\circ &\circ &\circ 
\\  \circ &\bullet &\circ &\circ &\circ &
\circ \end {array} \right),
\end{equation}
with coefficients, 

\begin{align*}
\ko_{00}^{(2)}=& b_1 b_{s^2} S \gamma_2 \left( \frac{1}{k_1^2} + \frac{1}{k_2^2} \right) + b_1^2 \left\{ \left[ \gamma_2 \left( 1 + F \right) + \beta_{12}\right] \left[\frac{1}{k_1^2} + \frac{1}{k_2^2} \right] + \frac{\mu \beta_{11}}{k_1 k_2} + \frac{F \beta_{6} + G \beta_{7} }{k_3^2}\right\}\\
\ko_{02}^{(2)}=& =\frac{f b_{s^2} S \gamma_2}{k_1^2} - b_1 f \left[  \frac{\gamma_2 \left( 1+F  \right) + \beta_{12}}{k_1^2} + \frac{G \gamma_2 k_2^2 }{k_1^2 k_3^2} + \frac{\beta_{11} \mu}{k_1 k_2} + \frac{\beta_{12}}{k_2^2} + \frac{F \beta_{6} + G (\beta_{7} + \gamma_2)}{k_3^2} \right]  \\
&+ b_1 \gamma_1 \left[ \frac{\beta_{14} }{k_1^2} + \frac{\mu \beta_{15} }{k_1 k_2} + \frac{\beta_{16}}{k_2^2} - \frac{G \beta_{19}}{k_3^2}\right] - b_1^2 \left[ \frac{\left(\beta_{9} + E \beta_{10} + \beta_{13}\right)}{k_1^2} + f \gamma_2 \left(\frac{1}{k_1^2} + \frac{1}{k_2^2} \right) \right] \\
\ko_{04}^{(2)}=& f^2 \gamma_2 \left[ \frac{b_1}{k_1^2} + \frac{G k_2^2}{k_1^2 k_3^2} \right] + \frac{b_1}{k_1^2} \left[ f \left( \beta_{9} + E \beta_{10} + \beta_{13} \right) - \beta_{17} \gamma_1 \right]\\
\ko_{11}^{(2)}=& \frac{b_{s^2} S \gamma_1^2}{k_1 k_2} + b_1 \left[ \beta_{15} \gamma_1 \mu \left(\frac{1}{k_1^2} + \frac{1}{k_2^2} \right) + \beta_{14} \gamma_1 \left( \frac{k_1}{k_2^3} + \frac{k_2}{k_1^3} \right) + \frac{2 \beta_{16} \gamma_1 + \gamma_1^2\left(1+F\right)}{k_1 k_2} \right.\\
&\left.- \frac{G \left(k_1^2 + k_2^2\right) \left(\beta_{19} \gamma_1  + 2 f \gamma_2\right)}{k_1 k_2 k_3^2} \right]\\
\ko_{13}^{(2)} =& f \gamma_1 \left[ - \frac{k_2 \beta_{14}}{k_1^3} - \frac{\beta_{15} \mu}{k_1^2} - \frac{\beta_{16}}{k_1 k_2} + G k_2 \frac{\left(\beta_{19} - \gamma_1\right)}{k_1 k_3^2}  \right] + b_1 \left[ -\gamma_1 \left( \frac{k_2 \beta_{17}}{k_1^3} + \frac{\beta_{18}}{k_1 k_2}\right) \right.  \\
&\left. + \frac{f}{k_1 k_2} \left(  \beta_{8} + 2 \beta_{9} + 2 E \beta_{10} - \gamma_1^2 \right) + 2 f^2 \gamma_2 \left( \frac{1}{k_1 k_2 }  + \frac{k_2}{k_1^3 }\right)\right] \\
\ko_{15}^{(2)}=& \frac{f k_2}{k_1^3} \left[ \gamma_1 \beta_{17} - \gamma_2 \right] \\
\ko_{20}^{(2)}=& - \frac{f b_{s^2} S \gamma_2}{k_2^2}  + b_1 \left[ \frac{\beta_{16} \gamma_1 - f \beta_{12}}{k_1^2} + \frac{\beta_{14} \gamma_1 - f \left[\beta_{12} + \gamma_2\left(1+F\right) \right]}{k_2^2} + \mu \frac{\left(- f \beta_{11} + \gamma_1 \beta_{15}\right)}{k_1 k_2}  \right. \\
&\left. - f\frac{F \beta_{6} + G \left(\beta_{7} + \gamma_2 \right)}{k_3^2} - \frac{f G k_1^2 \gamma_2}{k_2^2 k_3^2}\right] - b_1^2 \left[ \frac{\beta_{9} + E \beta_{10} + \beta_{13}}{k_2^2}  + f \gamma_2 \left(\frac{1}{k_1^2}+\frac{1}{k_2^2} \right)\right] \\
\ko_{22}^{(2)}=&  f \gamma_1 \left[ - \left(\beta_{14} + \beta_{16} \right)\left(\frac{1}{k_1^2}+\frac{1}{k_2^2}\right)  - \frac{2 \beta_{15} \mu}{k_1 k_2} + \frac{2 G \left(\beta_{19} - \gamma_1\right)}{k_3^2}\right] + f^2 \left[ \beta_{12}\left(\frac{1}{k_1^2}+\frac{1}{k_2^2} \right) \right. \\
&\left. + \frac{\beta_{11} \mu}{k_1 k_2} + \frac{F \beta_{6} + G \left(\beta_{7} + 2 \gamma_2\right)}{k_3^2}  \right] + b_1 \left[  \left(3 f^2 \gamma_2 -\beta_{18} \gamma_1 \right) \left(\frac{1}{k_1^2}+\frac{1}{k_2^2} \right) \right. \\
&\left. + f \frac{\left(k_1^2 + k_2^2\right) \left(\beta_{9} + E \beta_{10} + \beta_{13} - \gamma_1^2\right)}{k_1^2 + k_2^2}   \right]\\
\ko_{24}^{(2)}=& \frac{f}{k_1^2} \left[ \gamma_1 \left(\beta_{17} + \beta_{18}\right)+ f\left(- \beta_{9} - E \beta_{10} - \beta_{13} + \gamma_1^2\right) - 2 f^2 \gamma_2 \right]\\
\ko_{31}^{(2)}=& - f \gamma_1 \left[ \frac{k_1 \beta_{14}}{k_2^3} + \frac{\beta_{15} \mu}{k_2^2} + \frac{\beta_{16}}{k_1 k_2} - \frac{G k_1 \beta_{19}}{k_2 k_3^2} \right] + \left[- f \gamma_1^2 + 2 f^2 \gamma_2 \right] \frac{G k_1}{k_2 k_3^2} \\
&+ b_1 \left[  \frac{k_1}{k_2^3} \left( -\beta_{17} \gamma_1 + 2 f^2 \gamma_2 \right) + f \frac{\beta_{8}  + 2 \beta_{9} + 2 E \beta_{10} - \gamma_1^2 }{k_1 k_2}  - \frac{\beta_{18} \gamma_1^2 - 2 f^2 \gamma_2}{k_1 k_2} \right] \\
\ko_{33}^{(2)}=&\frac{f}{k_1 k_2} \left[ 2 \beta_{18} \gamma_1 + f \left( - \beta_{8} - 2 \beta_{9} - 2 E \beta_{10} + 2 \gamma_1^2 \right) - 2 f^2 \gamma_2 \right]\\
\ko_{40}^{(2)}=& f^2 \gamma_2 \frac{G k_1^2}{k_2^2 k_3^2} - b_1 \gamma_1 \frac{\beta_{17}}{k_2^2} + b_1 f \left[ \frac{\beta_{9} + E \beta_{10} + \beta_{13}}{k_2^2} \right] + b_1 f^2 \frac{\gamma_2}{k_2^2} \\
\ko_{42}^{(2)}=& f \gamma_1 \frac{\left( \beta_{17} + \beta_{18}\right)}{k_2^2} + \frac{f^2}{k_2^2} \left[ -\beta_{9} - E \beta_{10} \beta_{13} + \gamma_1^2 \right] - f^3 \frac{2 \gamma_2}{k_2^2}\\
\ko_{51}^{(2)}=& f \frac{k_1}{k_2^3} \left[ \beta_{17} \gamma_1 - f^2 \gamma_2  \right]  .
\end{align*}

\begin{equation}
\ko^{(3)}_{ab}= 
 \left( \begin {array}{cccccc} \circ &\bullet &\circ &\bullet 
&\circ &\bullet \\  \bullet &\circ &
\bullet &\circ &\bullet &\circ \\  \circ &
\bullet &\circ &\bullet &\circ &\circ \\  
\bullet &\circ &\bullet &\circ &\circ &\circ 
\\  \circ &\bullet &\circ &\circ &\circ &
\circ \\  \bullet &\circ &\circ &\circ &
\circ &\circ \end {array} \right), 
\end{equation}
with coefficients,

\begin{align*}
\ko_{01}^{(3)}=& \gamma_1 \gamma_2 \frac{b_{s^2} S}{k_1^2 k_2} + b_1 \gamma_1 \left[\frac{\beta_{12}\left(k_1^2 + k_2^2\right)}{k_1^2 k_2^3} + \frac{\beta_{11} \mu}{k_1 k_2^2} + \frac{F \beta_{6} + G \beta_{7}}{k_2 k_3^2} \right] + b_1 \gamma_2 \left[ \frac{\beta_{14} + \beta_{16}}{k_1^2 k_2} \right. \\
&\left. + \frac{\beta_{15}\mu}{k_1 k_2^2} + \frac{\left(k_2 \beta_{14} + k_1 \beta_{15} \mu\right)}{k_1^4} + \frac{\beta_{16}}{k_2^3} - G \beta_{19} \left( \frac{1}{k_2 k_3^2} + \frac{k_2}{k_1^2 k_3^2}\right)\right] + b_1 \gamma_1 \gamma_2 \frac{\left(1+F\right)}{k_1^2 k_2} - b_1^2 \frac{\left({\beta_{4}-\beta_{3}} + E \beta_{5}\right)}{k_1^2 k_2}\\
\ko_{03}^{(3)}=& - \frac{f \gamma_2}{k_1^2} \left[ \frac{k_2 \beta_{14}}{k_1^2} + \frac{\beta_{15}\mu}{k_1} + \frac{\beta_{16}}{k_2} \right] + f \gamma_2 G k_2 \frac{\left[ \beta_{19} - \gamma_1 \right]}{k_1^2 k_3^2} - b_1 \left[ \frac{k_2 \beta_{17} \gamma_2}{k_1^4}  + \left[ \gamma_2 \beta_{17} - f \left( {\beta_{4}-\beta_{3}} + E \beta_{5}\right) \right. \right. \\
&\left. \left. + \gamma_1 \left( \beta_{9} + E \beta_{10} + \beta_{13} \right) + f \gamma_1 \gamma_2 \right] \frac{1}{k_1^2 k_2}\right]\\
\ko_{05}^{(3)}=&  f \gamma_2 \frac{k_2 \beta_{17}}{k_1^4} \\
\ko_{10}^{(3)}=& \gamma_1 \gamma_2 \frac{b_{s^2} S}{k_1 k_2^2} + b_1 \gamma_1 \left[ \frac{\left(k_1^2 + k_2^2\right)\beta_{12}}{k_1^3 k_2^2} + \frac{F \beta_{6} + G \beta_{7}}{k_1 k_3^2} + \frac{\beta_{11}\mu}{k_1^2 k_2}  \right] + b_1 \gamma_2 \left[ \frac{\left(k_1^2 + k_2^2\right) \beta_{16}}{k_1^3 k_2^2} \right. \\
&\left. + \frac{\left( k_1^2 + k_2^2 \right) \left( k_1 \beta_{14} + k_2 \beta_{15} \mu \right) }{k_1^2 k_2^4}  - G \beta_{19} \frac{\left( k_1^2 + k_2^2 \right)}{k_1 k_2^2 k_3^2}  \right] + b_1 \gamma_1 \gamma_2 \frac{\left[ 1 + F\right]}{k_1 k_2^2} - b_1^2 \frac{\left[ {\beta_{4}-\beta_{3}} + E \beta_{5}\right]}{k_1 k_2^2} \\
\ko_{12}^{(3)}=& \gamma_1^2 \left[ \frac{\beta_{14}}{k_1^3} + \frac{\beta_{15} \mu}{k_1^2 k_2} + \frac{\beta_{16}}{k_1 k_2^2} - G \frac{\beta_{19}}{k_1 k_3^2} \right] + f \left[ - \gamma_1 \frac{\left( k_1^2 + k_2^2 \right) \beta_{12}}{k_1^3 k_2^2} - \gamma_1 \left( \frac{F \beta_{6} + G \left[ \beta_{7} + 3 \gamma_2 \right]}{k_1 k_3^2} - \frac{\beta_{11} \mu }{k_1^2 k_2 }  \right) \right. \\
&\left. + \gamma_2 \left( - \frac{\beta_{14}}{k_1 k_2^2} - \frac{\beta_{15} \mu }{k_1^2 k_2} - \frac{\beta_{16}}{k_1^3} + 3 \frac{G \gamma_2}{k_1 k_3^2} \right) \right] - b_1 \left[ \frac{\beta_{13} \gamma_1}{k_1^3} + \frac{-f \left({\beta_{4}-\beta_{3}} + E \beta_{5} \right) + \beta_{8} \gamma_1}{k_1 k_2^2 } \right. \\
&\left. + \frac{\gamma_1 \left(2 k_1^2 + k_2^2\right)}{k_1^3 k_2^2} \left( \beta_{9} + E \beta_{10} + f \gamma_2 \right) + \frac{\left(k_1^2 + k_2^2\right) \beta_{18} \gamma_2}{k_1^3 k_2^2}  \right]\\
\ko_{14}^{(3)}=& \gamma_1 \frac{- \beta_{17} \gamma_1 + f^2 \gamma_2}{k_1^3} + f \gamma_1 \frac{\left[ \beta_{9} + E \beta_{10} + \beta_{13} \right]}{k_1^3} + f \gamma_2\frac{\beta_{18}}{k_1^3}\\
\ko_{21}^{(3)}=& \gamma_1^2 \left[  \frac{\beta_{14}}{k_2^3} + \frac{\beta_{15} \mu }{k_1 k_2^2} + \frac{\beta_{16}}{k_1^2 k_2} - \frac{G \beta_{19}}{k_2 k_3^2} \right] + f \left[  -\frac{\left( k_1^2 + k_2^2 \right)\beta_{12} \gamma_1 }{k_1^2 k_2^3}  - \gamma_1 \frac{F \beta_{6}}{k_2 k_3^2} +  G \gamma_1 \frac{-\beta_{7} - 3 \gamma_2}{k_2 k_3^2} \right. \\
&\left.+ G \gamma_2 \frac{\beta_{19}}{k_2 k_3^2} - \frac{\mu}{k_1 k_2^2} \left( \gamma_1 \beta_{11} + \gamma_2 \beta_{15} \right) - \gamma_2 \left( \frac{\beta_{14}}{k_1^2 k_2} + \frac{\beta_{16}}{k_2^3} \right) \right] + b_1 \left[ f \frac{\left( {\beta_{4}-\beta_{3}} + E \beta_{5} \right)}{k_1^2 k_2} - \gamma_2 \frac{\left( k_1^2 + k_2^2\right) \beta_{18}}{k_1^2 k_2^3} \right. \\
&\left. - \gamma_1 \gamma_2 f \left( \frac{1}{k_2^3} + \frac{2}{k_1^2 k_2}\right) - \gamma_1 E \frac{\left( k_1^2 + 2 k_2^2\right) \beta_{10} }{k_1^2 k_2^3} - \gamma_1 \frac{\left(k_1^2 + 2 k_2^2 \right) \beta_{9}}{k_1^2 k_2^3} - \gamma_1 \left( \frac{\beta_{8}}{k_1^2 k_2} + \frac{\beta_{13}}{k_2^3} \right) \right]\\
\ko_{23}^{(3)}=& -\gamma_1^2 \frac{\beta_{18}}{k_1^2 k_2} + f \gamma_1 \frac{\left[ \beta_{8} + 3 \beta_{9} + 3 E \beta_{10} + \beta_{13}\right]}{k_1^2 k_2} + f \gamma_2 \frac{\left[ \beta_{17} + \beta_{18}\right]}{k_1 k_2^2} + f^2 \frac{\left[ -{\beta_{4}+\beta_{3}} - E \beta_{5} + 3 \gamma_1 \gamma_2 \right]}{k_1^2 k_2}\\
\ko_{30}^{(3)}=& -f \gamma_2 \left[ \frac{k_1 \beta_{14}}{k_2^4} + \frac{\beta_{15} \mu}{k_2^3} + \frac{\beta_{16}}{k_1 k_2^2} \right] + f \gamma_2 G \frac{k_1 \left(\beta_{19} - \gamma_1 \right)}{k_2^2 k_3^2} - b_1 \gamma_1 \left[ \frac{\beta_{9} + E \beta_{10} + \beta_{13}}{k_1 k_2^2}  \right] \\
&- b_1 \gamma_2 \frac{\left(k_1^2 + k_2^2\right) \beta_{17} }{k_1 k_2^4} + b_1 f \frac{\left({\beta_{4}-\beta_{3}} + E \beta_{5} - \gamma_1 \gamma_2 \right)}{k_1 k_2^2}\\
\ko_{32}^{(3)}=& - \gamma_1^2 \frac{\beta_{18}}{k_1 k_2^2} + f \gamma_1 \frac{\left(\beta_{8} + 3 \beta_{9} + 3 E \beta_{10} + \beta_{13} \right)}{k_1 k_2^2} + f \gamma_2 \frac{\left(\beta_{17} + \beta_{18} \right)}{k_1 k_2^2} - f^2 \frac{\left({\beta_{4}-\beta_{3}} + E \beta_{5} - 3 \gamma_1 \gamma_2 \right)}{k_1 k_2^2}\\
\ko_{41}^{(3)}=& - \gamma_1^2 \frac{\beta_{17}}{k_2^3} + f^2 \frac{\gamma_1 \gamma_2}{k_2^3} + f \gamma_1 \frac{\left(\beta_{9} + E \beta_{10} + \beta_{13} \right)}{k_2^3} + f \gamma_2 \frac{\beta_{18}}{k_2^3}\\
\ko_{50}^{(3)}=& f \gamma_2 \frac{k_1 \beta_{17}}{k_2^4}.
\end{align*}

\begin{equation}
	\ko^{(4)}_{ab}= 
 \left( \begin {array}{cccccc} \bullet &\circ &\bullet &\circ 
&\bullet &\circ \\  \circ &\bullet &\circ &
\bullet &\circ &\circ \\  \bullet &\circ &
\bullet &\circ &\circ &\circ \\  \circ &
\bullet &\circ &\circ &\circ &\circ \\  
\bullet &\circ &\circ &\circ &\circ &\circ 
\\  \circ &\circ &\circ &\circ &\circ &
\circ \end {array} \right), 
\end{equation}
with coefficients, 

\begin{align*}
\ko_{00}^{(4)}=& \gamma_2^2 \frac{b_{s^2} S}{k_1^2 k_2^2} + b_1 \gamma_2 \left[ F \frac{\left(k_1^2 + k_2^2 \right) \beta_{6}}{k_1^2 k_2^2 k_3^2} + G \frac{\left( k_1^2 + k_2^2 \right) \beta_{7}}{k_1^2 k_2^2 k_3^2} + \frac{\left( k_1^2 + k_2^2\right) \beta_{11} \mu }{k_1^3 k_2^3} + \frac{\left(k_1^2 + k_2^2 \right)^2 \beta_{12}}{k_1^4 k_2^4} \right] \\
&+ b_1 \gamma_2^2 \frac{\left( 1 + F \right)}{k_1^2 k_2^2} + b_1^2 \frac{\left( \beta_1 + E \beta_2 \right)}{k_1^2 k_2^2}\\
\ko_{02}^{(4)}=& \gamma_1 \gamma_2 \left[ \frac{\beta_{14}}{k_1^4} + \frac{\beta_{15} \mu}{k_1^3 k_2} + \frac{\beta_{16}}{k_1^2 k_2^2} - G \frac{\beta_{19}}{k_1^2 k_2^2} \right] - f \gamma_2 \left[ F \frac{\beta_{6}}{k_1^2 k_3^2} + G \frac{\beta_{7}}{k_1^2 k_3^2} + \frac{\beta_{11} \mu }{k_1^3 k_2} + \frac{ \left( k_1^2 + k_2^2 \right) \beta_{12}}{k_1^4 k_2^2} \right] \\
&- b_1 f \frac{\left( \beta_1 + E \beta_2 + \gamma_2^2 \right)}{k_1^2 k_2^2} - b_1 \gamma_1 \frac{\left( {\beta_{4}-\beta_{3}} + E \beta_{5} \right)}{k_1^2 k_2^2} - b_1 \gamma_2 \left[ \frac{\left( k_1^2 + k_2^2 \right) \beta_{9}}{k_1^4 k_2^2} + E \frac{\left( k_1^2 + k_2^2\right) \beta_{10} }{k_1^4 k_2^2} + \frac{\beta_{13}}{k_1^4} \right] \\
\ko_{04}^{(4)}=& - \gamma_1 \gamma_2 \frac{\beta_{17}}{k_1^4} + f \gamma_2 \frac{\left( \beta_{9} + E \beta_{10} + \beta_{13} \right)}{k_1^4}\\
\ko_{11}^{(4)}=& \gamma_1^2 \left[ F \frac{\beta_{6}}{k_1 k_2 k_3^2} + G \frac{\beta_{7}}{k_1 k_2 k_3^2} + \frac{\beta_{11} \mu}{k_1^2 k_2^2} + \frac{\left( k_1^2 + k_2^2\right) \beta_{12}}{k_1^3 k_2^3} \right] - \gamma_2^2 f \frac{2 G}{k_1 k_2 k_3^2} + \gamma_1 \gamma_2 \left[ \frac{\left( k_1^2 + k_2^2 \right) \beta_{14}}{k_1^3 k_2^3} \right. \\
&\left. + 2 \frac{\beta_{15} \mu}{k_1^2 k_2^2} + \frac{\left( k_1^2 + k_2^2 \right) \beta_{16}}{k_1^3 k_2^3} - 2 G \frac{\beta_{19}}{k_1 k_2 k_3^2} \right] - b_1 \gamma_1 \frac{\left( k_1^2 + k_2^2\right) \left( {\beta_{4}-\beta_{3}} + E \beta_{5} \right)}{k_1^3 k_2^3} - b_1 \gamma_2 \left[ \frac{\left( k_1^2 + k_2^2 \right) \beta_{8}}{k_1^3 k_2^3} \right. \\
&\left. + 2 \frac{\left( k_1^2 + k_2^2 \right) \beta_{9}}{k_1^3 k_2^3} + 2 E \frac{\left(k_1^2 + k_2^2 \right) \beta_{10}}{k_1^3 k_2^3} \right] - b_1 f \gamma_2^2 \frac{\left( k_1^2 + k_2^2 \right)}{k_1^3 k_2^3 } \\
\ko_{13}^{(4)}=& f^2 \frac{\gamma_2^2}{k_1^3 k_2} + f \gamma_1 \frac{\left( {\beta_{4}-\beta_{3}} + E \beta_{5} \right)}{k_1^3 k_2} + f \gamma_2 \frac{\left( \beta_{8} + 2 \beta_{9} + 2 E \beta_{10} \right)}{k_1^3 k_2} - \gamma_1^2 \frac{\left( \beta_{9} + E \beta_{10} + \beta_{13} \right)}{k_1^3 k_2} - \gamma_1 \gamma_2 \frac{\left( \beta_{17} + \beta_{18} \right)}{k_1^3 k_2} \\
\ko_{20}^{(4)}=& \gamma_1 \gamma_2 \left[ \frac{\beta_{14}}{k_2^4} + \frac{\beta_{15} \mu }{k_1 k_2^3} + \frac{\beta_{16}}{k_1^2 k_2^2} - G \frac{\beta_{19}}{k_2^2 k_3^2} \right] - f \gamma_2 \left[ F \frac{\beta_{6}}{k_2^2 k_3^2} + G \frac{\beta_{7}}{k_2^2 k_3^2} + \frac{\beta_{11} \mu}{k_1 k_2^3} + \frac{\left( k_1^2 + k_2^2 \right) \beta_{12}}{k_1^2 k_2^4} \right] \\
&- f \gamma_2^2 \frac{G}{k_2^2 k_3^2} - b_1 \gamma_1 \frac{\left( {\beta_{4}-\beta_{3}} + E \beta_{5} \right)}{k_1^2 k_2^2} - b_1 \gamma_2 \left[ \frac{\left( k_1^2 + k_2^2 \right) \beta_{9}}{k_1^2 k_2^4} + E \frac{\left( k_1^2 + k_2^2 \right) \beta_{10} }{k_1^2 k_2^4} + \frac{\left( k_1^2 + k_2^2 \right) \beta_{13}}{k_1^2 k_2^4} \right] \\
&+ b_1 f \frac{\left( \beta_1 - E \beta_2 - \gamma_2^2 \right)}{k_1^2 k_2^2} \\
\ko_{22}^{(4)}=& - \gamma_1^2 \frac{\left( \beta_{8} + 2 \beta_{9} + 2 E \beta_{10} \right)}{k_1^2 k_2^2} - \gamma_1 \gamma_2 \frac{2 \beta_{18}}{k_1^2 k_2^2}+  f \gamma_1 \frac{2\left( {\beta_{4}-\beta_{3}} + E \beta_{5} \right)}{k_1^2 k_2^2} + f \gamma_2 \frac{2 \left( \beta_{9} + E \beta_{10} + \beta_{13} \right)}{k_1^2 k_2^2} \\
&+ f^2 \frac{\left( \beta_1 + E \beta_2 + 2 \gamma_2^2 \right)}{k_1^2 k_2^2} \\
\ko_{31}^{(4)}=& f^2 \frac{\gamma_2^2}{k_1 k_2^3} + f \gamma_1 \frac{\left( {\beta_{4}-\beta_{3}} + E \beta_{5}\right)}{k_1 k_2^3} + f \gamma_2 \frac{\left( \beta_{8} + 2 \beta_{9} + 2 E \beta_{10}  \right)}{k_1 k_2^3} - \gamma_1^2 \frac{\left( \beta_{9} + E \beta_{10} + \beta_{13} \right)}{k_1 k_2^3} - \gamma_1 \gamma_2 \frac{\left( \beta_{17} + \beta_{18} \right)}{k_1 k_2^3} \\
\ko_{40}^{(4)}=&  - \gamma_1 \gamma_2 \frac{\beta_{17}}{k_2^4} + f \gamma_2 \frac{\left( \beta_{9} + E \beta_{10} + \beta_{13} \right)}{k_2^4}  .
\end{align*}
\begin{equation}
	\ko^{(5)}_{ab}= 
 \left( \begin {array}{cccccc} \circ &\bullet &\circ &\bullet 
&\circ &\circ \\  \bullet &\circ &\bullet &
\circ &\circ &\circ \\  \circ &\bullet &
\circ &\circ &\circ &\circ \\  \bullet &
\circ &\circ &\circ &\circ &\circ \\  
\circ &\circ &\circ &\circ &\circ &\circ 
\\  \circ &\circ &\circ &\circ &\circ &
\circ \end {array} \right), 
\end{equation}
with coefficients, 

\begin{align*}
\ko_{01}^{(5)}=& \gamma_1 \gamma_2 \left[ \frac{\left( F \beta_{6} + G \beta_{7} \right)}{k_1^2 k_2 k_3^2} + \frac{\beta_{11} \mu}{k_1^3 k_2^2} + \frac{\left( k_1^2 + k_2^2 \right) \beta_{12} }{k_1^4 k_2^3} \right] + \gamma_2^2 \left[ \frac{\beta_{14}}{k_1^4 k_2} + \frac{\beta_{15} \mu }{k_1^3 k_2^2} + \frac{\beta_{16}}{k_1^2 k_2^3} - G \frac{\beta_{19}}{k_1^2 k_2 k_3^2} \right] \\
&+ b_1 \gamma_1 \frac{\left( \beta_1 + E \beta_2 \right)}{k_1^2 k_2^3} - b_1 \gamma_2 \frac{\left( k_1^2 + k_2^2\right) \left( {\beta_{4}-\beta_{3}} + E \beta_{5} \right)}{k_1^4 k_2^3} \\
\ko_{03}^{(5)}=& - \gamma_1 \gamma_2 \frac{\left( \beta_{9} + E \beta_{10} + \beta_{13} \right)}{k_1^4 k_2} - \gamma_2^2 \frac{\beta_{17}}{k_1^4 k_2} + f \gamma_2 \frac{\left( {\beta_{4}-\beta_{3}} + E \beta_{5} \right)}{k_1^4 k_2 }\\
\ko_{10}^{(5)}=&\gamma_1 \gamma_2 \left[ \frac{\left( F \beta_{6} + G \beta_{7} \right)}{k_1 k_2^2 k_3^2} + \frac{\beta_{11} \mu}{k_1^2 k_2^3} + \frac{\left( k_1^2 + k_2^2 \right) \beta_{12}}{k_1^3 k_2^4}\right] + \gamma_2^2 \left[ \frac{\beta_{14}}{k_1 k_2^4} + \frac{\beta_{15} \mu}{k_1^2 k_2^3} + \frac{\beta_{16}}{k_1^3 k_2^2} - G \frac{\beta_{19}}{k_1 k_2^2 k_3^2} \right] \\
&+ b_1 \gamma_1 \frac{\left( \beta_1 + E \beta_2 \right)}{k_1^3 k_2^2} - b_1 \gamma_2 \frac{\left( k_1^2 + k_2^2\right) \left( {\beta_{4}-\beta_{3}} + E \beta_{5} \right)}{k_1^3 k_2^4} \\
\ko_{12}^{(5)}=& - \gamma_1^2 \frac{\left( {\beta_{4}-\beta_{3}} + E \beta_{5} \right)}{k_1^3 k_2^2} - \gamma_2^2 \frac{\beta_{18}}{k_1^3 k_2^2} - \gamma_1 \gamma_2 \frac{\left( \beta_{8} + 3 \beta_{9} + 3 E \beta_{10} + \beta_{13} \right) }{k_1^3 k_2^2} - f \gamma_1 \frac{\left( \beta_1 + E \beta_2 \right)}{k_1^3 k_2^2} + f \gamma_2 \frac{\left( {\beta_{4}-\beta_{3}} + E \beta_{5} \right)}{k_1^3 k_2^2}\\
\ko_{21}^{(5)}=& - \gamma_1^2 \frac{\left( {\beta_{4}-\beta_{3}} + E \beta_{5} \right)}{k_1^2 k_2^3} - \gamma_2^2 \frac{\beta_{18}}{k_1^2 k_2^3} - \gamma_1 \gamma_2 \frac{\left( \beta_{8} + 3 \beta_{9} + 3 E \beta_{10} + \beta_{13} \right)}{k_1^2 k_2^3} - f \gamma_1 \frac{\left( \beta_1 + E \beta_2 \right)}{k_1^2 k_2^3} + f \gamma_2 \frac{\left[ {\beta_{4}-\beta_{3}} + E \beta_{5} \right]}{k_1^2 k_2^3} \\
\ko_{30}^{(5)}=& -\gamma_1 \gamma_2 \frac{\left( \beta_{9} + E \beta_{10} + \beta_{13} \right)}{k_1 k_2^4} - \gamma_2^2 \frac{\beta_{17}}{k_1 k_2^4} + f \gamma_2 \frac{\left( {\beta_{4}-\beta_{3}} + E \beta_{5} \right)}{k_1 k_2^4} .
\end{align*}

\begin{equation}
	\ko^{(6)}_{ab} = 
 \left( \begin {array}{cccccc} \bullet &\circ &\bullet &\circ 
&\circ &\circ \\  \circ &\bullet &\circ &
\circ &\circ &\circ \\  \bullet &\circ &
\circ &\circ &\circ &\circ \\  \circ &
\circ &\circ &\circ &\circ &\circ \\  
\circ &\circ &\circ &\circ &\circ &\circ 
\\  \circ &\circ &\circ &\circ &\circ &
\circ \end {array} \right), 
\end{equation}

with coefficients, 

\begin{align*}
\ko_{00}^{(6)}=& \gamma_2^2 \left[ \frac{\left( F \beta_{6} + G \beta_{7} \right)}{k_1^2 k_2^2 k_3^2} + \frac{\beta_{11} \mu}{k_1^3 k_2^3} + \frac{\left( k_1^2 + k_2^2 \right) \beta_{12}}{k_1^4 k_2^4} \right] + b_1 \gamma_2 \frac{\left( k_1^2 + k_2^2\right) \left( \beta_1 + E \beta_2 \right)}{k_1^4 k_2^4}\\
\ko_{02}^{(6)}=& - \gamma_1 \gamma_2 \frac{\left( {\beta_{4}-\beta_{3}} + E \beta_{5} \right)}{k_1^4 k_2^2} - \gamma_2^2 \frac{\left( \beta_{9} + E \beta_{10} + \beta_{13} \right)}{k_1^4 k_2^2} - f \gamma_2 \frac{\left( \beta_1 + E \beta_2 \right)}{k_1^4 k_2^2} \\
\ko_{11}^{(6)}=& \gamma_1^2 \frac{\left( \beta_1 + E \beta_2 \right)}{k_1^3 k_2^3} - 2 \gamma_1 \gamma_2 \frac{{\beta_{4}-\beta_{3}} + E \beta_{5}}{k_1^3 k_2^3} - \gamma_2^2 \frac{\left( \beta_{8} + 2 \beta_{9} + 2 E \beta_{10} \right)}{k_1^3 k_2^3} \\
\ko_{20}^{(6)}=& - \gamma_1 \gamma_2 \frac{\left( {\beta_{4}-\beta_{3}} + E \beta_{5} \right)}{k_1^2 k_2^4} - \gamma_2^2 \frac{\left( \beta_{9} + E \beta_{10} + \beta_{13} \right)}{k_1^2 k_2^4} - f \gamma_2 \frac{\left( \beta_1 + E \beta_2 \right)}{k_1^2 k_2^4} .
\end{align*}

\begin{equation}
	\ko^{(7)}_{ab}= 
 \left( \begin {array}{cccccc} \circ &\bullet &\circ &\circ &
\circ &\circ \\  \bullet &\circ &\circ &
\circ &\circ &\circ \\  \circ &\circ &
\circ &\circ &\circ &\circ \\  \circ &
\circ &\circ &\circ &\circ &\circ \\  
\circ &\circ &\circ &\circ &\circ &\circ 
\\  \circ &\circ &\circ &\circ &\circ &
\circ \end {array} \right), 
\end{equation}

with coefficients, 
\begin{align*}
\ko_{01}^{(7)}=& \gamma_1 \gamma_2 \frac{\left( \beta_1 + E \beta_2 \right)}{k_1^4 k_2^3} - \gamma_2^2 \frac{\left( {\beta_{4}-\beta_{3}} + E \beta_{5}\right) }{k_1^4 k_2^3} \\
\ko_{10}^{(7)}=& \gamma_1 \gamma_2 \frac{\left( \beta_1 + E \beta_2 \right)}{k_1^3 k_2^4} - \gamma_2^2 \frac{\left( {\beta_{4}-\beta_{3}} + E \beta_{5} \right)}{k_1^3 k_2^4} .
\end{align*}

\begin{equation}
	\ko^{(8)}_{ab}= 
 \left( \begin {array}{cccccc} \bullet &\circ &\circ &\circ &
\circ &\circ \\  \circ &\circ &\circ &
\circ &\circ &\circ \\  \circ &\circ &
\circ &\circ &\circ &\circ \\  \circ &
\circ &\circ &\circ &\circ &\circ \\  
\circ &\circ &\circ &\circ &\circ &\circ 
\\  \circ &\circ &\circ &\circ &\circ &
\circ \end {array} \right) , 
\end{equation}
with coefficient

\begin{equation*}
	\ko_{00}^{(8)}= \gamma_2^2 \frac{\left( \beta_1 + E \beta_2 \right)}{k_1^4 k_2^4}.
\end{equation*}

\section{Squeezed limit}

These are the leading contributions to the squeezed limits for the multipoles~-- up to $O(\cH/k)$ and \(\ell = 3\).

\begin{align}
B_{0,0} &=-2 \sqrt{\pi} \frac{P_LP_S}{105} \left[f^{4}+\left(-3 b_{1}-\frac{15}{7}\right) f^{3}+\left(-77 b_{1}^{2}-33 b_{1}-14 b_{2}+\frac{14}{3} b_{s^2}\right) f^{2}-105\left(b_{1}^{2}+\frac{31}{21} b_{1}+\frac{4}{3} b_{2}\right.\right.\nonumber\\
&\left.\left.-\frac{4}{9} b_{s^2}\right) b_{1} f-195 b_{1}^{2}\left(b_{1}+\frac{14 b_{2}}{13}-\frac{14 b_{s^2}}{39}\right)\right]\\
B_{1,1}&=\sqrt{6 \pi} \frac{P_L P_S}{105 k_{L}} \left\{ \gamma_1 f^{3}+\left(18 b_{1} \gamma_1-9 \beta_{14}+6 \beta_{16}-5 \beta_{17}+2 \beta_{18}+\frac{15}{7} \gamma_1\right) f^{2}+\left[ \vphantom{\frac{\gamma^2}{3}} 49 \gamma_1 b_{1}^{2}+\left(-42 \beta_{14}\right.\right.\right.\nonumber \\
&\left.\left. \left. +56 \beta_{16}-24 \beta_{17}+12 \beta_{18}+18 \gamma_1\right) b_{1}+14 \gamma_1\left(b_{2}-\frac{b_{s^2}}{3}\right)\right] f-35\left[\left(\beta_{14}-2 \beta_{16}+\frac{3 \beta_{17}}{5}-\frac{2 \beta_{18}}{5}\right.\right.\right.\nonumber
\\
&\left.\left.\left.-\frac{13 \gamma_1}{7}\right) b_{1}-2 \gamma_1\left(b_{2}-\frac{b_{s^2}}{3}\right)  \right] b_{1}\right\} \\
B_{2,0} &=- 4 \sqrt{5 \pi}  f \frac{P_L P_S}{1155} \left[f^{3}+\left(-\frac{55}{14}-\frac{11 b_{1}}{2}\right) f^{2}+\left(-110 b_{1}^{2}-\frac{429}{7} b_{1}-11 b_{2}+\frac{11}{3} b_{s^2}\right) f\right.\nonumber\\
&\left. - 231 \frac{b_1}{2} \left(b_{1}^{2}+\frac{23}{21} b_{1}+\frac{2}{3} b_{2}-\frac{2}{9} b_{s^2}\right)  \right]\\
B_{2,2} &= 2 \sqrt{30 \pi} f  \frac{P_L P_S}{1155} \left[f^{3}+\left(-\frac{11 b_{1}}{6}-\frac{55}{42}\right) f^{2}+\left(-44 b_{1}^{2}-\frac{99}{7} b_{1}-11 b_{2}+\frac{11}{3} b_{s^2}\right) f\right. \nonumber\\
& \left.- 77 \frac{b_1}{2} \left(b_{1}^{2}+\frac{13}{7} b_{1}+2 b_{2}-\frac{2}{3} b_{s^2}\right) \right] \\
B_{3,1}&=\sqrt{21 \pi }\frac{P_L P_S}{165 k_{L}}   \left\{  \gamma_1 f^{3} +\left[\frac{418}{21} b_{1} \gamma_1-\frac{33}{7} \beta_{14}+\frac{22}{7} \beta_{16}-5 \beta_{17}+2 \beta_{18}+\frac{440}{147} \gamma_1\right] f^{2} + \left[ \frac{143 \gamma_1 b_{1}^{2}}{7}+ \right.\right. \nonumber \\
&\left.\left. \left( -\frac{99 \beta_{14}}{7}+\frac{22 \beta_{16}}{7}-\frac{77 \beta_{17}}{3}+\frac{242 \beta_{18}}{21}+\frac{792 \gamma_1}{49}\right) b_{1}+ \frac{88 \gamma_1}{7} \left( b_{2}-\frac{b_{s^2}}{3}\right) \right] f - b_1^2 \frac{132}{7} \left(\beta_{17}-\frac{2 \beta_{18}}{3}\right)  \right\}\\
B_{3,3} &=-f \sqrt{35 \pi}\frac{P_L P_S}{1155 k_{L}} \left[\gamma_1 f^{2}+\left(\frac{22 b_{1} \gamma_1}{3}-5 \beta_{17}+2 \beta_{18}-11 \beta_{14}+\frac{22 \beta_{16}}{3}\right) f-11\left(-3 b_{1} \gamma_1+\beta_{17}-\frac{2 \beta_{18}}{3}\right.\right.\nonumber\\
& \left.\left. \vphantom{\frac{b_{s^2}}{3}}+3 \beta_{14}-6 \beta_{16}\right) b_{1}\right]  
\end{align}

\acknowledgements CC was supported by the UK STFC (Consolidated Grant ST/P000592/1).  SJ and RM are supported by the South African Radio Astronomy Observatory (SARAO) and the National Research Foundation (Grant No. 75415). RM and OU are supported by the UK STFC (Consolidated Grant ST/N000668/1).

\bibliography{ref}

\end{document}